\documentclass[final,5p,times, twocolumn]{elsarticle}
\usepackage{amsmath,amssymb,amsfonts}
\usepackage{graphicx,color}
\usepackage{textcomp}
\usepackage{xcolor}
\usepackage{tcolorbox}
\usepackage{url}
\usepackage{hyperref}
\usepackage{placeins}
\usepackage{array}
\usepackage{booktabs}
\usepackage{float}
\usepackage{rotating}
\usepackage{multirow}
\usepackage{threeparttable}
\usepackage{xcolor}    % For the turn environment
\usepackage[T1]{fontenc}
\usepackage{titlesec}
\usepackage{makecell}
\titleformat{\paragraph}
{\normalfont\normalsize\itshape}{\theparagraph}{1em}{}
\titlespacing*{\paragraph}
{0pt}{3.25ex plus 1ex minus .2ex}{1.5ex plus .2ex}

\begin{document}
	\begin{frontmatter}
		
		\title{The Death Spiral of Open Source Projects: A Post-Mortem Analysis of Pull Request Workflow Dynamics}
		\author{Mohit Kaushik, Kuljit Kaur Chahal} %% Author name
		%% Author affiliation
		\affiliation{organization={Department of Computer Science, Guru Nanak Dev University},%Department and Organization
			addressline={}, 
			city={Amritsar},
			postcode={143005}, 
			state={Punjab},
			country={India}}
	\begin{abstract}
		Open Source Software projects (OSS) are central to modern technology, yet their survival rates remain low. Prior research has examined project mortality through macro-level indicators such as commit activity, developer abandonment, and ecosystem dependencies, but the micro-level dynamics of the Pull Request (PR) workflow have been largely overlooked. This study provides the first large-scale post-mortem analysis of PR workflows across 1,736 inactive GitHub repositories and 1.3 million human-driven PRs. Using a mixed-method quantitative design, we investigate three dimensions of mortality. First, our comparative descriptive analysis shows that workflow friction, extended review cycles, and negativity penalties are endemic properties of the entire GitHub platform across both active and inactive projects. Rejected PRs consistently attract higher discussion and negativity regardless of project health. Second, our evolutionary analysis identifies a universal ``death spiral" marked by declining innovation rates, exponential backlog growth, rising merge latency. The collapse was defined by silence and disengagement. Labeling formalization remained endemic throughout the lifecycle, while toxicity did not intensify. Finally, our explanatory modeling demonstrates that project lifespan is not determined by workflow efficiency but by inherent value and ecosystem dynamics. Popularity and innovation emerge as strong positive predictors of survival, while friction, rejection rates, labeling formalization, and negativity scale with longevity as byproducts rather than causes of failure. Robustness checks across alternative inactivity thresholds confirm these findings. Together, this work reframes OSS mortality as a socio-technical phenomenon in which abandonment and ecosystem value dominate survival outcomes, while PR-level workflow discipline plays a secondary role.
	\end{abstract}

	%% Keywords

	\end{frontmatter}
	
	% --- arXiv Acceptance Note ---
	\let\thefootnote\relax\footnotetext{\textbf{Accepted for publication in the Journal of Systems and Software.}}
	
	\section{Introduction}\label{sec:intro}
	
	Open-source software (OSS) has transformed the traditional way of building and distributing software~\cite{hoffmann2024value}. Numerous organizations use OSS to some extent~\cite{Lawson2023GlobalS2}. Some popular projects, such as Linux, Android, and the Apache ecosystem, show its widespread adoption and role in shaping the entire industry~\cite{hoffmann2024value}. While a community celebrates the success of OSS adoption, another community continuously highlights its failures. Previous studies report that 50\% of projects across major ecosystems npm, R, WordPress, and Laravel) eventually die~\cite{coelho2017modern}. This failure rate is not just limited to some ecosystems rather it is common for all projects~\cite{ait2022empirical}. However, the impact of this mortality is represented by critical vulnerability in the global software supply chain. When a project is abandoned, it creates a ``zombie dependency." It remains in use but no longer receives security patches. Recent research indicates that over 42\% of npm packages have not been updated in more than two years~\cite{pu2025assessing,reid2023large}. The consequences of this neglect were illustrated by the Log4j crisis, where a vulnerability compromised millions of systems globally. Furthermore, project mortality imposes a severe economic cost on downstream dependencies~\cite{linaaker2022characterize}. If the upstream fails or has vulnerabilities, these will trickle down to their downstream projects. Therefore, understanding the early micro‑signals of this decline is important for securing critical software infrastructure.  This contrast between the success stories and failure rate highlights a survivorship bias~\footnote{It is a tendency to focus on popular, ``surviving" projects while overlooking the vast number of projects that failed or were abandoned.}
	
	A significant number of studies have investigated the causes and predictors of project failure, often using post-mortem methodologies. Prior work has highlighted factors contributing to project abandonment, such as the loss of core developers~\cite{cosentino2015assessing,avelino2019abandonment}, abandonment of upstream dependencies~\cite{hasan2025understanding}, and maintainer burnout~\cite{coelho2017modern}. Other studies have applied survival analysis~\cite{robinson2022two}, machine learning prediction models~\cite{coelho2018identifying,liao2019prediction}, and ecosystem perspectives~\cite{Valiev2021}. Together, these studies suggest that project mortality is influenced by a combination of technical, social, and organizational factors.
	
	While these macro-level factors are important, the dynamics of day-to-day workflows have received far less attention~\cite{Golzadeh2019,Ortu2020}. Some research has examined workflow indicators to assess project health and to distinguish between active and dormant projects~\cite{Sbcars2024,Kikas2016}, but the emphasis has largely been on measuring activeness rather than modeling them as predictors of survival or decline. In particular, pull request (PR) dynamics have been overlooked in mortality research~\cite{Alami2020,zhang2022pull}. Within modern OSS ecosystems such as GitHub, the PR workflow serves as the central hub of innovation, governance, and community interaction~\cite{Wessel2023}. It is therefore the most logical place to observe early symptoms of decline, including stalled merges, expanding backlogs, community friction, and diminishing discussion~\cite{Asri2019,Sanei2021}. To the best of our knowledge, no prior study has conducted a post-mortem analysis of PR workflows to understand project mortality. We address this gap by analyzing 1,736 inactive OSS projects (comprising $\approx$1.3 million human-driven PRs) alongside a structurally matched control group of 1,736 active projects (comprising $\approx$2.67 million human-driven PRs). Together, this dataset of nearly 3.97 million pull requests and over 6.33 million comments allows us to isolate genuine signals of decline from platform-wide norms.
	
	In this study, we define \textit{Project Mortality} as the permanent end of development activity, including commits and merges. It is important to note that not all inactivity signals failure. As shown in our case studies (Section 5), inactivity can result from Strategic Completion, where a project fulfills its purpose and is archived (for example, Facebook Buck). It can also result from Abandonment, where contributors leave due to friction (for example, Simple-Gallery).
	
	While the intent behind these outcomes differs, the workflow signals are the same. Silence, backlog growth, and the end of maintenance all appear in the pull request record. For this reason, we use the term Mortality to cover the range of end‑states where a project stops functioning as an active software effort.
	
	\noindent The contributions of this study are:
	\begin{enumerate}
		\item \textbf{RQ1 (Descriptive):} We observed that process friction and negativity are platform-wide norms rather than unique markers of death. By comparing inactive projects against the active baseline, we found that rejected PRs consistently attract higher discussion volume and negative sentiment regardless of a project's overall health. Therefore, the presence of day-to-day workflow friction is a universal characteristic of open-source collaboration, rather than a direct cause of project mortality.
		
		\item \textbf{RQ2 (Evolutionary):} This study identified the universal mechanism of collapse, a ``death spiral" characterized by social abandonment, an exponential increase in backlog, declining innovation rates, and a resulting spike in merge friction. Labeling formalization remained endemic throughout the lifecycle, and toxicity did not intensify. Instead, collapse was marked by silence and disengagement.
		
		\item \textbf{RQ3 (Explanatory):} We demonstrated that lifespan is predicted by inherent project value and ecosystem dynamics rather than workflow efficiency. While popularity and innovation are strong predictors of lifespan, we also found an increasing prevalence of workflow issues, including labeling formalization, high rejection rates, and negative interactions, in aged projects. These dynamics appear as byproducts rather than causes of project failure. Robustness across different inactivity thresholds further confirms these findings.
	\end{enumerate}
	
	The remainder of this paper is organized as follows: Section~\ref{sec:LR} reviews related work on pull request workflows and project mortality. Section~\ref{sec:methodology} describes the methodology of the study. Section~\ref{sec:Results} then presents the results, followed by Section~\ref{sec:casestudy}, which aligns our findings with specific project examples. Section~\ref{sec:Discussion} interprets the findings, and Section~\ref{sec:Threats} discusses limitations and threats to validity. Finally, Section~\ref{sec:Conclusion} concludes the study and suggests directions for future research.

	\section{Literature Review}\label{sec:LR}
	
	OSS projects often face survival challenges. A project may exhibit early signs of decline that can be observed in day-to-day PR workflows. Prior works have explored these projects' decline indicators from different viewpoints, highlighting the necessity of synthesis. Therefore, this section is structured around two complementary perspectives. Section~\ref{subsec:LR1} discusses macro-level indicators at the project level, while Section~\ref{subsec:LR2} explains them from the micro-dynamics of pull requests. Section~\ref{subsec:LR3} discusses the role of automation or bot activities, and Section~\ref{subsec:LR_Gap} highlights the gap that our study addresses.
	
	\subsection{The ``Macro" View: Factors in OSS Project Mortality}\label{subsec:LR1}
	
	Based on our review, we categorize the macro factors into five different categories.
	
	\textbf{Socio-technical and Human Infrastructure:} As OSS projects are typically developed by a community of developers, the human infrastructure is a crucial component of a project's survival. This refers to the composition, behavior, organization, and stability of a community. Most OSS projects rely on a small number of core developers. Their abandonment or detachment is often measured by ``Truck Factor" or ``Developer Detachment" (TFDD). Analyzing the TFDD, Avelino et al.~\cite{avelino2019abandonment} found that 65\% of projects had a TF of 1 or 2. This indicates the project's fragility and dependence on a very small number of developers. Similarly, Nourry et al.~\cite{nourry2024myth} analyzed over 36,000 projects and noted that 89\% of projects experienced at least one TFDD. The loss of these core developers contributes to a project's decline. They also reported that only 27\% of projects survived a TFDD by attracting new core developers. The survival also becomes difficult if a TFDD occurs early, specifically within the first three years of a project's life~\cite{nourry2024myth}. In addition to TFDD, team size also impacts the project survival. However, it is often debated that merely the sheer volume of contributors does not guarantee survival, rather the type of contributors matters more~\cite{robinson2022two}. Studies have proven the ``enough eyeballs" hypothesis and found that larger teams increase survival. Robinson et al.~\cite{robinson2022two} found that projects with more than 20 unique contributors had a higher survival probability compared to small teams. Ait et al.~\cite{ait2022empirical} further confirm that projects with a larger community size (Tier 3) have a higher chance of survival. The team size also matters in innovative projects. Fang et al.~\cite{Fang2024} found that the innovative projects, while attracting more popularity, tend to have smaller teams and face survival challenges.
	
	Conversely, Song and Kim~\cite{Song2018OSS} argue that a simple increase in contributors can be ``poisonous" to efficiency due to communication overhead. Their results showed that ``heavy contributors" positively affect the project efficiency, whereas a simple increase in total contributors negatively affects it. In addition to these attributes, developers' own experience with the project and their states also influence the project survival~\cite{kaur2022exploring}. Calefato et al.~\cite{Calefato2022Inactivity} modeled developer inactivity using ``sleeping" and ``dead" metaphors. They found that all core developers take breaks (sleeping state), but about 45\% completely disengage (dead state) for at least two years.
	
	Recent studies highlight that governance, culture, and non-coding labor are as important as human infrastructure for a project's survival. Lenge et al.~\cite{lange2025invisible} define invisible labor as unrecognized work such as mentorship, conflict resolution, and administrative tasks. They argue that this labor is the ``backbone" of OSS. Therefore, a lack of leadership skills in these areas affects the newcomers' onboarding and ultimately project survival. Additionally, the formal and informal rules governing a project significantly affect its survival. Yin et al.~\cite{Yin2021,Yin2022} used institutional analysis and revealed that projects that exhibit a higher frequency of institutional statements survive longer. Additional factors such as mentorship, governance, and ``volunteer's dilemma" also impact the project survival~\cite{Yin2022,Miller2023}.
	
	\textbf{Activity, Stability, and Rhythm Indicators:} These indicators serve as ``vital signs" of a project. Beyond simple aggregate counts, these indicators analyze the cadence, consistency, and responsiveness of project activities. Researchers often use them to distinguish between active (healthy) and dormant (unhealthy) projects. These indicators include, but are not limited to, commits, releases, and issue resolution. Adejumo et al.~\cite{adejumo2025commit} utilize the composite stability index and highlight that the commit volume alone is insufficient. They revealed that projects with maintaining stable commit patterns are less risky and more resilient. Other studies also found that projects with high revision frequency and a low ratio of non-working days have a slightly higher probability of survival~\cite{robinson2022two,liao2019prediction}. Additionally, the release dynamics also impact the survival. The release speed and interval are significant predictors of survival. Hasan et al.~\cite{hasan2025understanding} analyzed the Maven ecosystem's libraries and observed a transition from fast to slow in release speeds, followed by inactivity. Although this pattern is not always a sign of abandonment, some libraries exhibit a spike of high frequency before decline.
	
	Survival does not solely depend on just adding a new code, but on the capacity to maintain it. Park and Kwon~\cite{park2025analyzing} define and analyze the issue retention rate (IRR) as a proxy for maintenance capacity. The authors observed that as projects age or grow in code complexity, the IRR tends to rise, increasing the risk of abandonment. With a similar focus on maintainer-centric features, Xu et al.~\cite{xu2025predicting} revealed that maintainer responsiveness and delay are critical factors of survival. Additionally, recent findings indicate that sustained active interaction (issues and comments) drives extreme longevity, whereas the predictive power of passive metrics declines over time~\cite{kaushik2026community,kaushik2026beyond}.
	
	\textbf{Code and Innovation:} These indicators investigate the ``technical health" of a project to determine its mortality. Project growth is generally seen as a survival signal. Khondu et al.~\cite{khondhu2013all} found that active projects tend to launch with a larger initial codebase and grow consistently than dormant or inactive projects. Liao et al.~\cite{liao2019prediction} further supported this by reporting a strong positive Pearson correlation (r = 0.85) between project size and its lifespan. Additionally, code quality and maintainability also have an impact on survival. However, this relationship is complex. Khondu et al.~\cite{khondhu2013all} investigated the maintainability index (MI) and revealed that the majority of inactive and abandoned projects have a stable MI. Furthermore, recent studies focused on innovation revealed a tension between project novelty and sustainability. Fang et al.~\cite{Fang2024} noted a higher risk of abandonment in innovative projects due to a limited pool of maintenance labor. Similarly, Xu et al.~\cite{xu2025predicting} introduced the feature ratio as a proportion of innovation and bug fix PRs. They identified a shrinking feature ratio as a marker of ``functional stagnation," which serves as an early  warning signal for project abandonment.
	
	\textbf{Ecosystem Signals:} Beyond the internal code activity patterns, these indicators represent user engagement and governance practices. Modeling ``user-repository" networks, He et al.~\cite{he2024revealing} proposed a metric called Repository Centrality derived from the Hyperlink-Induced Topic Search (HITS) algorithm. They revealed that a drop in HITS weight indicates a decline in repository prevalence. Similarly, Xu et al.~\cite{xu2025predicting} analyzed user-centric features' contribution in their survival study. They noted that projects that retain influential users' engagement are more robust against abandonment. Other engagement metrics, such as stars and watchers, also contribute to a project's survival~\cite{park2025analyzing,he2024revealing}. Additionally, survival is heavily influenced by governance type, structure, and policies. Yin et al.~\cite{Yin2022} analyzed outside (managerial) and inside (reflexive or community-driven) governance and noted that successful projects demonstrated active self-governance. Likewise, the organizational structure also influences the project abandonment risk. Ait, Coleho et al.~\cite{ait2022empirical,coelho2020github} both found a higher survival probability of organizations' owned projects. Other factors like resource funding also influence the project decline.
	
	\textbf{Project's Macro-Demographics:} A project's own attributes, such as project age, programming language, and application domain, influence its trajectory. Nourry et al.~\cite{nourry2024myth} highlighted maturity as a buffer and revealed that a mature project survives the TFDD event where the core development team leaves. Avelino et al.~\cite{avelino2019abandonment} challenged this view and highlighted maturity as rigidity. They noted that surviving projects were younger at the time of TFDD compared to non-surviving projects. This view is further supported by Park and Kwon~\cite{park2025analyzing} from the IRR perspective. Besides project age, the choice of programming language also affects its survival. Liao et al.~\cite{liao2019prediction} noted  significant differences in average lifespan across various languages. Coleho et al.~\cite{coelho2020github} further supported this and found a higher survival probability for Ruby projects. Additionally, the project's application domain is also important. Adjeumo et al.~\cite{adejumo2025commit} found that blockchain projects are more stable compared to front-end or UI frameworks. Other application domains, such as system software and web libraries, also had a higher survival probability.
	
	While these studies provide a robust understanding of high-level mortality factors, they often lack detailed analysis of the specific, day-to-day contribution workflow.
	
	\subsection{The ``Micro" View: PR Dynamics as Health Indicators}\label{subsec:LR2}
	
	Some authors have focused on the ``micro-level" dynamics of the Pull Request (PR) workflow. While, these studies are essential for establishing metrics of project health, but they focus mainly on \textit{active, healthy projects}, which introduces a ``survivorship bias.''
	
	In modern collaborative software development, the PR is the central mechanism of contribution, and the choice to accept or reject is an important, socio-technical decision~\cite{Alami2020}. This decision is influenced by PR's technical quality, adherence to guidelines, and contribution complexity~\cite{Alami2020,zhang2022pull}, all of which can contribute to friction during the review process. For our mortality analysis, we are concerned with measuring the impact of this friction.
	
	\textit{Discussion volume} serves as a key proxy for this friction. Prior studies indicate a complex pattern: rejected PRs tend to generate more comments and longer discussion times (median of 1.69 days compared to 0.6 days for accepted PRs)~\cite{Golzadeh2019}, while high interaction intensity can be associated with accepted ones. This high discussion volume directly indicates complexity or disagreement, often leading to PR abandonment and wasted effort. Research confirms that abandoned PRs tend to be more complex with longer review times, and the number of responses is a significant predictor of abandonment~\cite{Golzadeh2019}.
	
	Discussion volume alone, however, is insufficient. The emotions or sentiment present within the discussion also significantly influence the PR decision. \textit{Sentiment analysis} is increasingly used to understand community interactions \cite{Ortu2020,Sanei2021,Asri2019}. Positive sentiment encourages collaboration and faster reviews. Conversely, negative sentiments like frustration or anger can affect progress and escalate conflicts. Asri et al.~\cite{Asri2019} found that the reviews with positive sentiment close 1.32 days faster than those with negative sentiment. The strong negative emotions like anger or dominance correlate with a lower probability of a PR being merged~\cite{Sanei2021, Ortu2020}. A strong correlation also exists between sentiment and the final outcome. Positive sentiments were found in 91.81\% of successful reviews, compared to 64.44\% in aborted reviews~\cite{Asri2019}. Together, these studies highlight sentiment as a key indicator of \textit{community health} and a potential predictor of maintainer burnout.
	
	Finally, PR labels are a key organizational tool for managing this workflow~\cite{Sbcars2024}. By categorizing PRs (e.g., bug, feature), maintainers direct attention and streamline the process. While the presence of labels can impact latency~\cite{Kikas2016}, their absence is also a critical signal. A high proportion of unlabeled PRs has traditionally been interpreted as a potential breakdown in project governance and process maturity~\cite{Alami2020}, which are known factors in long-term sustainability.
	
	\subsection{The Role of Automation and Bot Activity}\label{subsec:LR3}
	
	The socio-technical dynamics of PRs are further complicated by the pervasive role of automation. Studies highlights that bots are now integral to modern OSS development~\cite{wessel2021don}. Bots are widely used to automate tasks such as dependency updates (e.g., Dependabot), code reviews (e.g., Codecov), and backlog management by closing stale issues and PRs (e.g., Stale bot)~\cite{wessel2022quality,khatoonabadi2023understanding}.
	
	This bot activity fundamentally differs from human interaction. Bots often generate the first response, sometimes within seconds, which artificially shortens the `time-to-first-response' metric~\cite{hasan2023understanding}. However, this automated reply does not accelerate human engagement; in fact, studies show that the first human response is often significantly slower in PRs that receive a bot-first response~\cite{hasan2023understanding}. Furthermore, automated comments, which often follow repetitive patterns, do not reflect the project's `social health' and can be perceived by developers as noise or a distraction~\cite{golzadeh2021ground,wessel2021don}.
	
	Automated processes can, therefore, distort the metrics used to measure project health. Labels such as `auto-merge,' when applied by bots, may reduce merge times but can also artificially inflate signals of project activity~\cite{wessel2022quality}.
	
	\subsection{The Research Gap}\label{subsec:LR_Gap}
	
	A significant gap exists at the intersection of these research fields. Most direct post-mortem or large-scale survival studies (as discussed in Sections~\ref{subsec:LR1} and~\ref{subsec:LR2}) rely heavily on commit activity, release patterns, and issue resolution as primary failure metrics. These studies often overlook the detailed Pull Request (PR) workflow dynamics for mortal analysis.
	
	In recent years, more advanced predictive models have begun to fill this gap by incorporating PR metrics as features to forecast decline. For instance:
	\begin{itemize}
		\item Predictive models for \textit{Level of Maintenance Activity (LMA)} use PR counts (open, closed, merged) as input features~\cite{coelho2020github}.
		\item Lifecycle classification models use \textit{PR\_review\_duration\_in\_hours} and \textit{PR\_average\_commits} to distinguish mature, ``graduated'' projects from earlier stages~\cite{lu2025open,lumbard2024empirical}.
		\item Abandonment prediction models use \textit{avg\_response\_time} to issues and PRs, as well as the \textit{feature\_ratio} (proportion of feature-related PRs), to detect ``functional stagnation'' and maintainer disengagement~\cite{xu2025predicting}.
	\end{itemize}
	
	In essence, while PR metrics are increasingly used as features for forecasting decline, to our knowledge, no study has yet conducted a granular, retrospective post-mortem analysis with PR workflow dynamics (such as \textit{merge friction}, \textit{social abandonment}, and \textit{governance collapse}) as the central subject of an investigation into inactive projects.
	
	Our study addresses this gap by providing the first in-depth ``autopsy''\footnote{Borrowing from the medical field, we use the term ``autopsy" metaphorically to denote a post-mortem analysis of failed OSS projects, examining their historical data to determine the causes of their `mortality'.} of the PR workflow and benchmarking these dynamics against a control group of active projects to isolate genuine signals of mortality.
	
	\section{Research Design and Methodology}\label{sec:methodology}
	
	This study employs a multi-stage, mixed-method quantitative design. We first conduct a descriptive and evolutionary analysis (RQ1, RQ2) to identify the ``symptoms of mortality" in inactive projects. We then use an explanatory regression model (RQ3) to identify the statistical predictors of total lifespan.
	
	\subsection{Research Questions and Hypotheses}\label{subsec:RQs}
	
	Our research is guided by three Research Questions, each associated with a set of testable hypotheses. This structure flows from a descriptive (RQ1), to an evolutionary (RQ2), to an explanatory (RQ3) analysis of project mortality.
	
	\subsubsection{RQ1: Descriptive Dynamics of PR Friction}\label{subsubsec:RQ1}
	
	Our first research question seeks to understand the ``micro-level'' characteristics of friction and work type.

		\noindent\textbf{RQ1:} What are the foundational characteristics of PR-level dynamics (discussion, labels, sentiment) and their associations with PR-level outcomes (merge time, merge likelihood, closure) in inactive projects?

	To answer this, we test four hypotheses based on our literature review (Section~\ref{sec:LR}):
	
	\begin{itemize}
		\item \textbf{H1a (Work Type):} Innovation (enhancement) PRs have a significantly longer merge time than Maintenance (bugfix) PRs.
		
		\textit{Rationale:} Prior studies show that feature-oriented contributions introduce higher complexity and dependencies, requiring extended review cycles, whereas bug fixes have clearer success criteria and are merged more quickly~\cite{Fang2024,xu2025predicting}.
		
		\item \textbf{H1b (Failure Cost):} Closed (rejected) PRs have a significantly higher discussion volume than Merged PRs.
		
		\textit{Rationale:} Rejected contributions often involve prolonged debate and coordination delays, consuming scarce review bandwidth and leading to higher discussion volume~\cite{iaffaldano2019developers,qiao2024code}.
		
		\item \textbf{H1c (Failure Toxicity):} Closed PRs are significantly more likely to contain at least one negative comment than Merged PRs.
		
		\textit{Rationale:} Negative emotions such as frustration or anger correlate with lower merge probability and maintainer burnout, making closed PRs more likely to contain negative comments ~\cite{Calefato2022Inactivity,Asri2019}.
		
		\item \textbf{H1d (Toxicity Volume):} Closed PRs exhibit a significantly higher proportion of negative sentiment overall compared to Merged PRs.
		
		\textit{Rationale:} Sentiment analyses confirm that toxicity scales disproportionately in failed contributions, undermining collaboration and increasing the likelihood of rejection ~\cite{Asri2019,linaaker2024sustaining}.
	\end{itemize}

	\subsubsection{RQ2: Evolutionary Post-Mortem of Workflow}\label{subsubsec:RQ2}
	
	Our second research question investigates the ``symptoms of mortality'' by tracking our key metrics over the project lifecycle.

	\noindent\textbf{RQ2:} As these inactive projects approached their final commit, how did their PR workflow patterns (e.g., median merge time, rejection rate, discussion sentiment, governance maturity) evolve across their lifecycle?

	To investigate the ``Death Spiral'' pattern, we test the following hypotheses by comparing the first quartile (Q1) to the final quartile (Q4):
	
	\begin{itemize}
		\item \textbf{H2a (Stagnation):} The innovation rate is significantly lower in Q4 compared to Q1.
		
		\textit{Rationale:} As projects approach abandonment, they exhibit functional stagnation, shifting from generative labor (adding new features) to reactive maintenance. This is reflected in a shrinking ``Feature Ratio"~\cite{xu2025predicting}, the long-term risk of sustaining highly innovative projects~\cite{Fang2024}, and slowing release speeds preceding abandonment~\cite{hasan2025understanding}.

		\item \textbf{H2b (Friction Spike):} The median merge time is significantly higher in Q4 compared to Q1.
		
		\textit{Rationale:} Declining maintenance capacity leads to longer delays in processing contributions. Response latency is a signal of disengagement~\cite{xu2025predicting}, while rising issue retention rates in older projects reflect accumulated technical debt~\cite{park2025analyzing}.
		
		\item \textbf{H2c (Process Collapse):} The rejection rate is significantly higher in Q4 compared to Q1.

			\textit{Rationale:} Limited resources and weak governance increase rejection rates in later stages. Maintainers struggle with low-quality contributions~\cite{linaaker2024sustaining}, and failed projects often lack guidelines or integration practices that raise barriers to merging ~\cite{coelho2020github}.

		\item \textbf{H2d (Backlog Explosion):} The number of open PRs are significantly higher at the end of Q4 compared to Q1.

			\textit{Rationale:} Mortality is marked by an inability to process incoming requests, creating a growing backlog. Issue retention reflects failing capacity~\cite{park2025analyzing}, while unresolved dependencies accumulate in a volunteer's dilemma~\cite{Miller2023}.

		\item \textbf{H2e (Social Collapse):} The median discussion volume is significantly lower in Q4 compared to Q1.

			\textit{Rationale:} Collapse is often preceded by social silence. The ``dead" state is defined by absence of communication signals~\cite{iaffaldano2019developers,Calefato2022Inactivity}, while lack of interest manifests in reduced interaction~\cite{coelho2017modern}.

		\item \textbf{H2f (Toxicity Spike):} The proportion of PRs containing negative sentiment are significantly higher in Q4 compared to Q1.

			\textit{Rationale:} Decline often coincides with rising social friction and burnout. Core developers express stronger negative sentiment linked to inactivity~\cite{kaur2022exploring}, while user aggression places additional pressure on maintainers~\cite{linaaker2024sustaining}.

		\item \textbf{H2g (Labeling formalization):} The proportion of unlabeled PRs is significantly higher in Q4 compared to Q1.

			\textit{Rationale:} The absence of organizational housekeeping signals governance failure. Sustainable projects exhibit active self-governance~\cite{Yin2022}, while failed projects often lack templates and guidelines, leaving contributions untriaged ~\cite{coelho2017modern}.

	\end{itemize}

	\subsubsection{RQ3: Explanatory Modeling of Lifespan}\label{subsubsec:RQ3}
	
	Our final research question moves from observation to explanation, seeking to identify the statistical predictors of total lifespan.
	
		\noindent \textbf{RQ3:} Which aggregated PR workflow attributes are the strongest statistical predictors of a project's total lifespan, after controlling for programming language, license, and project size?
	
	To answer this, we test the following explanatory hypotheses, derived from prior literature linking workflow friction, labeling formalization, innovation, and community sentiment to project lifespan:
	
	\begin{itemize}
		\item \textbf{H3a (Friction):} A higher median merge time is a significant negative predictor of lifespan.
		  
		\textit{Rationale:} Longer merge times reflect response delays and accumulated technical debt, both of which weaken project sustainability~\cite{xu2025predicting,park2025analyzing}.
		
		\item \textbf{H3b (Waste):} A higher proportion of closed PRs is a significant negative predictor of lifespan.
		  
		\textit{Rationale:} A high rate of rejected PRs represents wasted effort and discourages contributors, adding pressure on maintainers and reducing survival chances~\cite{linaaker2024sustaining,coelho2017modern}.
		
		\item \textbf{H3c (Labeling formalization):} A higher proportion of unlabeled PRs is a significant negative predictor of lifespan. 
		 
		\textit{Rationale:} Missing labels indicate weak governance and poor workflow organization, which are linked to project decline~\cite{Yin2022,coelho2020github}.
		
		\item \textbf{H3d (Toxicity):} A higher average proportion of negative sentiment is a significant negative predictor of lifespan.
		  
		\textit{Rationale:} Negative sentiment erodes collaboration and increases the risk of disengagement, contributing to abandonment~\cite{linaaker2024sustaining,iaffaldano2019developers}.
		
		\item \textbf{H3e (Innovation):} A higher proportion of innovation PRs is a significant positive predictor of lifespan. 
		 
		\textit{Rationale:} Innovation signals ongoing relevance and attracts community interest, which supports survival, while stagnation is a marker of decline~\cite{xu2025predicting,Fang2024}.
	\end{itemize}
	
	\subsection{Data Collection \& Sample Definition}\label{subsec:data_collection}
	
	\subsubsection{Data Source}
	
	To conduct the ``post-mortem" analysis, we selected the dataset provided by~\cite{Dabic2021}, accessible via the SEART platform\footnote{\url{https://seart-ghs.si.usi.ch/}}. This platform hosts data of over 1.7 million repositories. Initially, we filtered the projects with a minimum of 50 contributors, resulting in a dataset of 34{,}972 repositories. This threshold was adopted to exclude toy or personal projects and to ensure the selection of projects with a substantive collaborative history~\cite{Mastropaolo2023_ICSE}. The observational window for this study covers the entire history of these projects up to the data collection cutoff on September 30, 2024. This ensures that our analysis captures long-term evolutionary trends rather than short-term fluctuations.
	
	\subsubsection{Inactive Project Sample Definition} 
	
	We applied a multi-stage filtering process to isolate the suitable projects for this analysis. First, we excluded fork repositories, as forks often inherit the code, issues, and other properties from their upstream source and might have different activity patterns or lifespan~\cite{Pietri2020}. Also, some forks might be used for experimentation or backup purposes. Therefore, including them in the analysis can lead to inaccurate conclusions.
	
	Second, we selected the projects that contain at least 10 pull requests. This threshold aligns with other empirical studies. Researchers often use these thresholds to avoid misleading results. For example, Dey and Mockus applied a similar threshold of at least 5 pull requests~\cite{dey2020effect}, while Qiao et al. excluded pull requests with sparse histories in their analysis~\cite{qiao2024code}. Other studies also support the idea of excluding projects with limited pull request activity~\cite{thongtanunam2016reviewer}.
	
	Finally, to categorize the project as inactive, we applied a six-month threshold on the last commit date. Aligning with the project lifespan definition~\cite{liao2019prediction}, the projects with no commits in the six months with respect to the data collection date were marked as inactive. Although, using a single threshold for defining the project state is often debated and lacks consensus~\cite{Calefato2022Inactivity,evangelopoulos2008determining}. We explicitly address the arbitrary nature of this threshold in sensitivity analysis (Section~\ref{subsec:sensitivity_analysis}). 
	
	The filtering process was as follows (Table~\ref{tab:filter}):
	
	\begin{itemize}
		\item \textbf{Step 1 – Metadata Completeness:} 
		We removed 19,556 repositories that lacked the metadata required for this study: 12,487 with missing \texttt{createdAt} (partly null or set to the default date 01/01/1970, preventing reliable lifespan calculation), 
		3,214 with missing \texttt{isFork} (preventing fork exclusion), 
		and 3,855 with missing \texttt{pullRequests} (preventing activity verification).

		\item \textbf{Step 2 – Activity Threshold:}  
		We removed 1,804 repositories with fewer than 10 pull requests to exclude toy projects.
		
		\item \textbf{Step 3 – Fork Exclusion:}  
		We removed 307 forked repositories to avoid inheriting signals from upstream projects.
		
		\item \textbf{Step 4 – Inactivity Classification:}  
		We retained 1,736 inactive projects that had no commits in the six months prior to the data collection date (September 30, 2024). This step excluded 11,689 active repositories.
	\end{itemize}

	This process yielded our final cohort of 1,736 inactive repositories. From this cohort, we used the \textit{PyGithub}\footnote{\url{https://pypi.org/project/PyGithub/}} library to retrieve all associated Pull Requests, resulting in a final dataset of 1,473,360 PRs for analysis.
	
	\begin{table}[!h]
		\centering
		\caption{Repository Selection and Reduction Summary}
		\label{tab:filter}
		\begin{tabular}{@{}ll@{}}
			\toprule
			\textbf{Step} & \textbf{Repos Remaining} \\ \midrule
			Initial dataset & 34,972 \\
			Drop missing values & 15,416 \\
			Fewer than 10 PRs & 13,612 \\
			Remove forks & 13,305 \\
			Keep inactive only & 1,736 \\ \bottomrule
		\end{tabular}
	\end{table}
	
	Control Group Selection: To address potential survivorship bias and establish a robust baseline, we constructed a 1:1 structurally matched control group of active projects from the repositories excluded in Step 4. From 11,689 active projects, we selected 1,736 active projects to mirror the size of our inactive cohort, with a 95\% confidence level, a 2\% margin of error, and a 50\% population proportion to maximize variance. To ensure comparability, we verified that the control group shared similar structural attributes with the inactive cohort, summarized in Table~\ref{tab:struct}
	
	\begin{table}[!h]
		\centering
		\caption{Structural comparison between Inactive and Active (Control) cohorts}
		\label{tab:struct}
			\begin{tabular}{lcc}
				\toprule
				\textbf{Attribute} & \textbf{Inactive Cohort} & \textbf{Active Cohort} \\
				\midrule
				Median Popularity   & 1,404 stars   & 1,588 stars   \\
				Median Size         & 13,438 KB     & 23,714 KB     \\
				PL1  & JavaScript (21.5\%) & Python (17.4\%) \\
				PL2  & Python (13.5\%)     & TypeScript (11.9\%) \\
				PL3  & TypeScript (9.4\%)  & Go (10.4\%) \\
				\bottomrule
			\end{tabular}
		\vspace{0.7em}
		\begin{minipage}{0.9\linewidth}
			\footnotesize \textit{Note: PL1, PL2, PL3 denote the top three primary programming languages by proportion.}
		\end{minipage}
	\end{table}

	 Both groups demonstrate a heavy concentration of modern web and infrastructure languages and share comparable median popularity and repository size. This alignment confirms that our control group represents successful survivors of a similar scale and domain. As the SEART platform hosts project metadata, we mined the full history of these 1,736 active repositories. This resulted in a massive initial dataset of 3,284,608 pull requests and 5,567,614 associated comments. To align with the inactive data timeline and avoid look-ahead bias, we restricted data mining to September 30, 2024.
	
	\subsection{Data Preparation and Operationalization}\label{subsec:data_prep}
	
	To prepare our dataset for analysis, we performed a multi-stage operationalization process to filter noise, classify contributions, and derive our key predictor variables.
	
	\subsubsection{Bot Filtering and Preprocessing}
	
	As discussed in our literature review (Section~\ref{subsec:LR3}), bot activity fundamentally differs from human interaction and can distort socio-technical metrics~\cite{hasan2023understanding, wessel2021don}. Therefore, to isolate the human signals of project mortality, we performed a two-stage filtering process on the 1,473,360 PR dataset:
	
	\begin{enumerate}
		\item We first removed all PRs where the author's login contained the string \texttt{bot}, a common heuristic for identifying automated accounts.
		\item We then removed all remaining PRs that contained bot-related labels (e.g., \texttt{auto\_merge}, \texttt{renovate}, \texttt{dependabot}), as these represent automated workflows.
	\end{enumerate}
	
	Since simple string matching can occasionally yield false positives (for example, legitimate accounts with names containing ``robotics" or ``botany"), we introduced an additional validation step. Flagged accounts were examined against their activity profiles. Those displaying repetitive behaviors such as dependency updates, automated merges, or continuous CI-related comments were confirmed as bots. In contrast, accounts showing a broader range of contributions, including code commits, documentation edits, and participation in discussions, were reinstated as human contributors. This combined manual and heuristic validation ensured that the filtering process excluded only genuine automated activity, while preserving authentic human-driven interactions.
	
	 While recent tools like RABBIT~\cite{RABBIT} or BotHunter~\cite{abdellatif2022bothunter} offer automated detection, we opted for a high-precision string-matching heuristic combined with manual validation. In socio-technical analysis, false positives (classifying a human as a bot) are more damaging than false negatives, as they remove genuine social signals. However, we also acknowledge that false negatives also pose a threat to validity, as they can incorrectly inflate human interaction metrics. Our approach prioritized preserving human interaction to reduce the risk of false positives.
	
	This filtering process yielded a final dataset of 1,296,100 human-driven PRs and 2,160,722 comments, which serves as the basis for all subsequent analyses. We also normalized all PR status fields (e.g., to ``Merged'' or ``Closed'') and converted timestamps to datetime objects for duration calculations. The same bot detection approach was applied to the control group. This process resulted in a final dataset containing 2,671,639 human-driven pull requests and 4,173,241 comments.
	
	\subsubsection{Work Type Classification (Label Analysis)}
	
	To test our hypotheses regarding ``innovation'' (H2a) and ``governance'' (H3c), we classified every PR into one of five mutually exclusive categories. This step moves beyond raw label counts and provides a defensible proxy for workflow maturity.
	
	We defined keyword families for three primary categories:
	\begin{itemize}
		\item \textbf{Innovation:} Labels indicating new features, performance improvements, or major refactoring (i.e., \texttt{feature}, \texttt{enhancement}, \texttt{performance}).
		\item \textbf{Maintenance:} Labels for bug fixes or dependency updates (i.e., \texttt{bug}, \texttt{fix}, \texttt{deps}).
		\item \textbf{Administration:} Labels for project upkeep (i.e.,\texttt{docs}, \texttt{ci}, \texttt{test}, \texttt{chore}).
	\end{itemize}
	
	 PRs with no labels were classified as unlabeled. We treat this as a proxy for on-platform labeling formalization. We acknowledge that many successful projects rely on external tools or implicit norms rather than GitHub labels. However, within the context of GitHub workflow data, the absence of labels represents a lack of explicit categorization, which correlates with the informal processes often seen in smaller or dying communities~\cite{Alami2020,Chakraborti2024Governance}. For labeled PRs, we applied a fixed priority order (Innovation $\rightarrow$ Maintenance $\rightarrow$ Administration) when multiple categories were present. This ordering preserves innovation signals, while still capturing maintenance and administrative works. Finally, PRs with labels outside these families (e.g., \texttt{help wanted}, \texttt{priority:high}) were classified as \textit{other\_labeled}. Overall, this procedure helps us to distinguish between labeling formalization, value-adding contributions, and other organizational signals.
	
	\noindent
	For explanatory modeling (RQ3), we operationalized work type as the proportion of innovation PRs (\texttt{prop\_innovation}) within each project. In RQ3, this variable is explicitly treated as a substantial predictor (H3e). This attribute measures the relative emphasis on value-adding contributions compared to maintenance and administrative work. By modeling innovation proportion directly, we highlight its theoretical relevance as a signal of project value and sustainability, and test its role as a positive determinant of lifespan.
	
	\subsubsection{Sentiment Analysis}
	
	To operationalize social friction, we analyzed the sentiment of all PR discussions. We employed a fine-tuned DistilBERT model specifically trained and validated on human-authored OSS discourse, as detailed in our prior work~\cite{kaushik2026sentiment}. The model is publicly available\footnote{\url{https://huggingface.co/iamohitkaushik1/distilbert-active-learning-github-sentiment}}.
	
	The training process began with a labeled seed dataset of 7{,}122 GitHub PR and commit comments (2,013 positive, 3,022 neutral, 2,087 negative), all written by human contributors, split into 70\% training, 15\% validation, and 15\% test sets. Initial fine-tuning achieved 92.07\% accuracy and an F1-score of 0.920. We then applied a hybrid active learning strategy combining prediction entropy and breaking-ties margin, iteratively expanding the labeled set until convergence was confirmed using the OracleAcc-MCS stopping criterion. The final model was evaluated on a held-out test set of 6,818 samples, achieving 96\% accuracy with precision, recall, and F1 all above 0.95, demonstrating robust generalization.
	
	For large-scale inference, the converged model was applied to 5.3 million preprocessed GitHub comments, all filtered to exclude automated bot activity. Within this corpus, our current study's dataset of 2,160,722 PR comments (drawn from inactive projects) forms a direct subset, ensuring that the sentiment predictions reported here are fully consistent with the large-scale inference results. The overall sentiment distribution was predominantly neutral (66.4\%), followed by positive (22.2\%) and negative (11.4\%), with prediction confidence consistently above 0.95.
	 
	\paragraph{Aggregation.} 
	We aggregated comment-level predictions into PR-level and project-level metrics in two steps. First, for each PR, we calculated the proportion of comments classified as negative, neutral, or positive (\texttt{prop\_negative}, \texttt{prop\_neutral}, \texttt{prop\_positive}). Each PR was then assigned a \textit{dominant sentiment} category based on the highest proportion. In rare cases of exact ties (e.g., equal positive and negative comments), we applied a two-step resolution strategy. If negative and positive sentiments were equal, we marked the PR as neutral, since both signals cancel each other out. If ties occurred with neutral, we favored the emotional category to preserve the signal of community sentiment. However, given that most PRs had only a single comment, such ties were negligible. Second, we averaged these PR-level proportions across all PRs in a project to obtain project-level metrics (e.g., \texttt{avg\_prop\_negative}), which serve as quantitative proxies for community tone and interaction quality across the lifecycle of inactive projects.
	
	\subsubsection{Lifecycle Quartile Definition (RQ2)}
	For RQ2, we divided each project's lifespan into four quartiles to capture the temporal evolution of workflow dynamics. Quartiles were defined by partitioning PR creation timestamps within each project into four equal calendar segments (Q1--Q4). Thus, each quartile represents one-fourth of the elapsed collaborative time. This temporal-based partitioning is methodologically crucial, as it allows us to observe periods of low-activity gridlock (e.g., few PRs taking a long time in Q4), which would be invisible under volume-based quartiles that artificially equalize contribution counts. In this way, the uneven distribution of PRs across quartiles is itself informative of decline dynamics and was retained as part of the analysis. We selected a quartile-based comparison (Q1 vs. Q4) rather than a median split (First Half vs. Second Half) to isolate the ``initial observable phase'' and ``final termination'' phases. A median split would average the long ``Maintenance Phase" (Q2 and Q3) into the results. This might dilute the sharp signals of decline (such as the backlog explosion) that is typically observed only in the final stage of the lifecycle.
	
	For each quartile, we aggregated median merge time, proportion of closed PRs, proportion of innovation PRs, and PR counts. At the project level (RQ3), we additionally computed the proportion of unlabeled PRs (used as a proxy for labeling formalization), sentiment proportions (negative, neutral, positive), and overall innovation share. Median values were used for time-based metrics to reduce sensitivity to extreme outliers. Projects with missing values (e.g., no PRs in a given quartile) were retained with \texttt{NaN} entries, which were handled appropriately in regression models to avoid bias.
	
	\subsection{Analysis Methods}\label{subsec:analysis_methods}
	
	We employed a specific statistical test for each research question, as our data (e.g., merge time, discussion volume) is not normally distributed. For all comparisons, we report both statistical significance ($p$-value) and a non-parametric effect size (Cliff's Delta, $\delta$). Following Romano et al.~\cite{romano2006appropriate}, we interpret $\delta$ values as small ($<0.147$), medium ($\approx0.33$), and large ($>0.474$), which provides a practical measure of effect magnitude in large datasets where $p$-values are often trivially small.
	
	\subsubsection{RQ1: Descriptive Dynamics}
	
	To test H1a--H1d, we applied non-parametric tests on the human-driven PR dataset:
	\begin{itemize}
		\item \textbf{Work Type (H1a):} Kruskal-Wallis H-test to assess differences in \texttt{human\_merge\_time\_hours} across \texttt{work\_type} categories.
		\item \textbf{Failure Cost and Toxicity Volume (H1b, H1d):} Mann-Whitney U test to compare distributions of \texttt{discussion\_volume} and volume-weighted negativity between merged and closed PRs.
		\item \textbf{Failure Toxicity Presence (H1c):} Chi-Square ($\chi^2$) test on a 2x2 contingency table to evaluate whether closed PRs were more likely to contain negative sentiment than merged PRs.
	\end{itemize}
	
	\subsubsection{RQ2: Evolutionary Post-Mortem}
	
	To test H2a--H2e, we analyzed aggregated quartile data. Because metrics are compared across four lifecycle stages within the same projects (e.g., Q1 vs.\ Q4), this constitutes paired data. We therefore used the Wilcoxon signed-rank test, a paired non-parametric method suitable for within-subject comparisons. Because we tested multiple hypotheses (H2a to H2g) over the same quartile populations, we applied a Bonferroni correction, adjusting our significance threshold to $\alpha =0.0071$ to control for family-wise error rates. Descriptive statistics of total counts (e.g., open PRs per quartile) were also reported to contextualize these trends.
	
	\subsubsection{RQ3: Explanatory Modeling}
	
	To test H3a--H3e, we employed a multi-stage regression framework to model project lifespan:
	
	\begin{itemize}
		\item \textbf{Model:} We specified an Ordinary Least Squares (OLS) multivariate regression with \texttt{lifeSpan} as the dependent variable and predictors including \texttt{median\_merge\_time}, \texttt{prop\_closed}, \texttt{prop\_unlabeled}, \texttt{avg\_prop\_negative}, \texttt{prop\_innovation}, \texttt{stargazers}, \texttt{language\_category}, and \texttt{size} (project scale). Innovation was operationalized as the continuous proportion of innovation PRs (\texttt{prop\_innovation}) within each project and treated as a substantive predictor (H3e). While survival models (e.g., Cox proportional hazards, Accelerated Failure Time) are standard for time-to-event data, they require censored observations to estimate hazard functions. Our study design is explicitly post-mortem: we analyze only inactive (i.e., completed) projects, with no right-censoring. In this context, OLS provides an appropriate and interpretable framework for modeling lifespan as a continuous outcome. Prior to estimation, we conducted diagnostic checks for skewness, kurtosis, and heteroscedasticity. Variables exhibiting extreme skew were transformed as needed (e.g., log-transformed) to stabilize variance and improve model fit. Sensitivity checks assess predictor stability across alternative inactivity thresholds.
		
		\item \textbf{Residual Robustness:} Because socio-technical data often violate normality assumptions, we planned to conduct residual diagnostics (Omnibus, Jarque-Bera, skewness, kurtosis). Anticipating non-normal residuals, we specified the use of heteroscedasticity-robust HC3 standard errors to ensure valid inference.
		
		\item \textbf{Dynamic Effects:} To capture heterogeneity across project types, we planned Quantile Regression at the 25th, 50th (median), and 75th percentiles of lifespan. This allows us to examine how predictors behave differently in fragile versus resilient projects, and to assess whether innovation (H3e) shows stronger effects in longer-lived projects.
	\end{itemize}

	\section{Results}\label{sec:Results}
	
	\subsection{Baseline Comparison: Active vs. Inactive Cohorts}\label{subsec:4.1}

\begin{table}[!h]
	\centering
	\caption{Comparison of PR Workflow Dynamics between Inactive (Dead) and Active (Control) Projects.}
	\label{tab:control_comparison}
		\begin{tabular}{l c c}
			\toprule
			Metric & Inactive  & Active\\
			\midrule
			Labeling Formalization & 72.7\% & 64.2\% \\
			Friction & 14.1 h & 16.8 h \\
			Rejection Rate & 26.3\% & 17.2\% \\
			Toxicity & 2.9\% & 6.8\% \\
			Discussion Volume & 1.67 & 1.52 \\
			\bottomrule
		\end{tabular}
	\vspace{0.5em}
	\begin{minipage}{0.9\linewidth}
		\footnotesize \textit{Note: Labeling Formalization is measured by the proportion of unlabeled PRs. 
			Friction values are median merge times (Inactive IQR = 81.2 h, Active IQR = 88.7 h). 
			Toxicity refers to negative sentiment proportions. 
			Discussion Volume values are mean comments per PR (Inactive SD = 3.63, Active SD = 3.96).}
	\end{minipage}
\end{table}

From Table~\ref{tab:control_comparison}, we found that labeling formalization is not a special marker of dying projects. High rates of unlabeled PRs are common across the platform, and both inactive and active groups show this pattern. Friction also does not uniquely signal collapse. In fact, active projects take slightly longer to merge (16.8 hours vs. 14.1 hours), showing that slower review is consistent with survival. Rejection, however, is different. Inactive projects close far more PRs, which means wasted contributor effort strongly correlates with death. Toxicity also does not drive collapse; negative sentiment is actually higher in active projects (6.8\% vs. 2.9\%). Discussion density provides additional context. Active projects average 1.52 comments per PR compared to 1.67 in inactive ones, with substantial standard deviations in both groups (3.96 and 3.63, respectively) indicating a heavy-tailed distribution where most PRs receive minimal attention. This indicates that a low overall volume of discussion is simply a baseline characteristic of the platform rather than a symptom of poor health. We therefore treat the sudden drop to silence (investigated further in RQ2) as the true signal of project death, rather than the overall baseline.

\subsection{Descriptive Statistics (RQ1 Foundation)}\label{subsec:DSRQ1}
	
	Having established in previous section that workflow friction is not the unique driver of project-level mortality, we now turn to RQ1 to test how this friction manifested in day-to-day operations. While friction may not kill the project, hypotheses H1a to H1d test whether it acted as a barrier to individual contributions. We begin by summarizing the overall distribution of pull requests (PRs) across our entire dataset. After filtering bot activity, the dataset contained 1,296,093 human-driven PRs from the 1,736 inactive projects and 2,671,639 human-driven PRs from the 1,736 active control projects.
	Table~\ref{tab:pr_distribution} reports the distribution of PRs across work types and outcomes for both cohorts. Applying our active control group baseline to these micro-level workflow dynamics reveals differences in integration efficiency. Active projects exhibit a similarly high proportion of unlabeled PRs (64.2\% vs. 72.7\% in inactive). This pattern confirms that low labeling formalization is a platform-wide norm rather than a unique symptom of decline. Regarding merge rate, active projects successfully merge 81.3\% of all PRs, compared to only 68.5\% in the inactive cohort. Furthermore, inactive projects suffer from a significantly higher rejection rate (26.3\% vs. 17.2\% for active) and a bloated backlog of open PRs at the time of data collection (5.2\% vs. 1.6\% for active). This indicates that while the types of work submitted to both cohorts are structurally similar, inactive projects uniquely struggle to integrate them successfully.
	
	\begin{table}[!h]
		\centering
		\caption{Distribution of Pull Requests by Work Type and Outcome in Inactive vs Active Projects (Proportions)}
		\label{tab:pr_distribution}
			\begin{tabular}{l r r}
				\toprule
				\multirow{2}{*}{\textbf{Category}} & \textbf{Inactive} & \textbf{Active} \\
				& (N = 1,296,093) & (N = 2,671,639) \\
				\midrule
				\multicolumn{3}{l}{\textit{Work Type}} \\
				Labeling Formalization & 72.7\% & 64.2\% \\
				Other Labeled          & 16.3\% & 19.6\% \\
				Administration         & 4.8\%  & 6.6\% \\
				Innovation             & 3.7\%  & 5.0\% \\
				Maintenance            & 2.5\%  & 4.6\% \\
				\midrule
				\multicolumn{3}{l}{\textit{PR Outcome}} \\
				Merged                 & 68.5\% & 81.3\% \\
				Closed (Rejected)      & 26.3\% & 17.2\% \\
				Open                   & 5.2\%  & 1.6\% \\
				\bottomrule
			\end{tabular}
			\vspace{0.5em}
			\begin{minipage}{0.9\linewidth}
				\footnotesize \textit{Note: Proportions are calculated relative to total PRs in each cohort (Inactive: 1,296,093; Active: 2,671,639).}
			\end{minipage}
	\end{table}
	
	Beyond overall distributions, outcomes varied sharply across work types (Table~\ref{tab:worktype_outcome}). By comparing these outcomes against the active baseline, we observe that the underlying hierarchy of integration success remains consistent across both cohorts. Administrative and Maintenance tasks achieve the highest merge rates, while Innovation and Labeling Formalization (unlabeled PRs) face higher baseline friction. However, the absolute integration efficiency between the cohorts differs drastically. Active projects maintain exceptionally high merge rates (ranging from 79.1\% to 86.9\%) and minimal open backlogs (1.1\% to 2.9\%) regardless of the work type. In contrast, inactive projects struggle significantly across the board, and their integration failure is most pronounced in unlabeled PRs. Inactive projects reject 26.7\% of unlabeled PRs (compared to 17.7\% in active) and leave 5.4\% PRs open at the time of death (compared to just 1.3\% in active). Even for explicitly labeled Innovative and Maintenance work, inactive projects reject these contributions at higher rates and leave a larger fraction unresolved in the backlog. This indicates that the elevated friction observed in inactive projects is not caused by the specific nature of the work being submitted, but rather by a systemic inability of the dying community to effectively review, integrate, and clear the queue.
	
	\begin{table*}[!h]
		\centering
		\caption{PR Outcomes by Work Type in Inactive vs Active Projects (Proportions)}
		\label{tab:worktype_outcome}
		\begin{tabular}{lccc|ccc}
			\toprule
			\multirow{2}{*}{\textbf{Work Type}} & \multicolumn{3}{c}{\textbf{Inactive} (N = 1,296,093)} & \multicolumn{3}{c}{\textbf{Active} (N = 2,671,639)} \\
			\cmidrule(lr){2-4} \cmidrule(lr){5-7}
			& Merged & Closed & Open & Merged & Closed & Open \\
			\midrule
			Administration         & 85.3\% & 13.2\% & 1.4\% & 86.9\% & 11.6\% & 1.5\% \\
			Labeling Formalization & 67.9\% & 26.7\% & 5.4\% & 80.9\% & 17.7\% & 1.3\% \\
			Innovation             & 79.9\% & 16.8\% & 3.2\% & 82.4\% & 14.7\% & 2.9\% \\
			Maintenance            & 81.3\% & 16.5\% & 2.2\% & 86.2\% & 12.8\% & 1.1\% \\
			Other Labeled          & 73.1\% & 21.2\% & 5.7\% & 79.1\% & 18.8\% & 2.1\% \\
			\bottomrule
		\end{tabular}
		\vspace{0.5em}
		\begin{minipage}{0.9\linewidth}
			\centering
			\footnotesize \textit{``Open'' refers to PRs that remained open at project death (inactive projects) and at the data collection date (active projects), respectively.}
		\end{minipage}
	\end{table*}
	
	We next examine workflow metrics and sentiment proportions (Table~\ref{tab:workflow_metrics}). Sentiment proportions were calculated at the comment level across all PRs, so values in Table~\ref{tab:workflow_metrics} do not sum to 100\%. When comparing these metrics against the active baseline, we observe that baseline workflow friction and minimal discussion are simply normal characteristics of the GitHub platform, rather than unique symptoms of mortality. On average, inactive projects actually exhibited faster merge times (14.1 hours) and closure times (53.4 hours) compared to active projects (16.8 hours and 90.4 hours, respectively). It suggests that slower, more deliberate reviews may be a sign of rigorous maintenance rather than dysfunction. Furthermore, median discussion volume is minimal across both cohorts, indicating that low collaborative deliberation on typical PRs is standard behavior. Importantly, the sentiment distribution is nearly identical between the two groups. Both cohorts exhibit exactly 2.9\% negative sentiment at the comment level, and neutral sentiment heavily dominates both environments. This highlights that the presence of baseline negativity and workflow friction does not differentiate dying projects from successful ones.
	
	\begin{table}[!h]
		\centering
		\caption{Median Workflow Metrics and Sentiment Distribution in Inactive vs Active Projects}
		\label{tab:workflow_metrics}
		\begin{tabular}{lrr}
			\toprule
			\textbf{Metric} & \textbf{Inactive} & \textbf{Active}\\
			\midrule
			Median Merge Time & 14.1 h & 16.8 h \\
			Median Closed Time & 53.4 h & 90.4 h \\
			Median Discussion Volume & 1.0 & 0.0 \\
			\midrule
			Negative Sentiment & 2.9\% & 2.9\% \\
			Neutral Sentiment  & 26.7\% & 27.4\% \\
			Positive Sentiment & 15.0\% & 11.5\% \\
			\bottomrule
		\end{tabular}
	\end{table}
	
	Sentiment also varied with PR outcomes (Table~\ref{tab:sentiment_outcome}). For categorical analysis, each PR was assigned a dominant sentiment based on the highest proportion of comments; in rare cases of exact ties (e.g., equal positive and negative comments), assignment defaulted to one category, though such cases were negligible given that most PRs had only a single comment. By comparing these distributions against the active baseline, we observe that the penalty for negative sentiment is a universal platform norm rather than a unique symptom of mortality. In both cohorts, positive sentiment overwhelmingly supports integration (over 81\% merged), while negative sentiment PRs face disproportionately high rejection rates (36.4\% in inactive vs. 33.4\% in active). Because active projects penalize toxicity just as harshly as inactive ones, we can conclude that friction and negative interactions are standard features of the GitHub review process. The true diagnostic difference between the cohorts appears in PRs with ``No Comments.'' In active projects, silence typically indicates frictionless, routine integration, resulting in the highest merge rate (86.0\%) and a minimal open backlog (1.3\%). In contrast, silence in inactive projects corresponds to neglect and abandonment. The merge rate for un-commented PRs drops to 69.6\%, and 7.1\% remain unresolved in the backlog at the time of project death. These patterns highlighting that while workflow friction is universal, the inability to process silent contributions is a unique marker of a dying project.

	\begin{table*}[!h]
		\centering
		\caption{PR Outcomes by Dominant Sentiment Category in Inactive vs Active Projects}
		\label{tab:sentiment_outcome}
		\begin{tabular}{lccc|ccc}
			\toprule
			\multirow{2}{*}{\textbf{Sentiment}} & \multicolumn{3}{c}{\textbf{Inactive} (N = 1,296,093) } & \multicolumn{3}{c}{\textbf{Active (N = 2,671,639)}} \\
			\cmidrule(lr){2-4} \cmidrule(lr){5-7}
			& Merged & Closed & Open & Merged & Closed & Open \\
			\midrule
			Positive    & 81.7\% & 16.5\% & 1.8\% & 84.3\% & 14.4\% & 1.4\% \\
			Neutral     & 68.2\% & 28.7\% & 3.1\% & 73.1\% & 24.8\% & 2.1\% \\
			Negative    & 60.8\% & 36.4\% & 2.8\% & 64.2\% & 33.4\% & 2.4\% \\
			No Comments & 69.6\% & 23.4\% & 7.1\% & 86.0\% & 12.8\% & 1.3\% \\
			\bottomrule
		\end{tabular}
	\end{table*}
	
	Having established categorical distributions, we now turn to workflow dynamics. Table~\ref{tab:resolution_times} reports the full distributional properties of merge and closed times for both cohorts. Both outcomes exhibited extreme skew and heavy tails across both active and inactive projects, confirming that highly skewed resolution times are a universal platform norm rather than a unique symptom of dysfunction. For merged PRs, the median resolution time in the inactive cohort was 14.1 hours, whereas the active cohort took slightly longer at 16.8 hours. In both groups, means were heavily inflated (227 hours and 194 hours, respectively) due to a small number of PRs that remained open for months or years. Skewness and kurtosis confirm that while most merges were rapid, some PRs experienced extreme delays regardless of project health. A similar baseline consistency appears for closed PRs. In both cohorts, the median resolution time (53.4 hours for inactive, 90.4 hours for active) was substantially longer than merges, suggesting that rejected contributions lingered before being dismissed. The mean closed time was consistently high (around 1,800 hours for both groups), with a maximum exceeding 105,000 hours ($\approx$12 years). Although skew and kurtosis were lower than for merges, they still indicate a heavy-tailed distribution where most rejections occurred within days but some PRs remained unresolved for years. Overall, these comparative distributions confirm our earlier observation that slower, more deliberate reviews do not indicate project decay, as successful active projects actually take significantly longer to formally close out rejected PRs than dying projects do.
	
	\begin{table*}[!h]
		\centering
		\caption{Descriptive statistics of PR resolution times in Inactive vs Active Projects}
		\label{tab:resolution_times}
		\begin{tabular}{lrr|rr}
			\toprule
			\multirow{2}{*}{\textbf{Statistic}} & \multicolumn{2}{c}{\textbf{Inactive} (N = 1,296,093)} & \multicolumn{2}{c}{\textbf{Active} (N = 2,671,639)} \\
			\cmidrule(lr){2-3} \cmidrule(lr){4-5}
			& Merged (hours) & Closed (hours) & Merged (hours) & Closed (hours) \\
			\midrule
			Mean      & 227.10   & 1804.70  & 193.94   & 1818.48 \\
			Std       & 1175.98  & 5395.18  & 942.47   & 5084.13 \\
			Median    & 14.10    & 53.37    & 16.83    & 90.42 \\
			Min       & 0.00     & 0.00     & 0.00     & 0.00 \\
			Max       & 79,547.41 & 102,880.66 & 75,042.97 & 105,942.79 \\
			Skew      & 15.45    & 5.59     & 17.55    & 5.51 \\
			Kurtosis  & 395.59   & 41.65    & 511.87   & 42.64 \\
			\bottomrule
		\end{tabular}
	\end{table*}
	
	Finally, we examine the discussion volume across outcomes (Table~\ref{tab:discussion_by_status}). Comparing these distributions against the active baseline reveals a divergence in how pending work is handled, while simultaneously confirming that debate over rejected contributions is a universal norm. In both cohorts, merged PRs were typically processed with minimal attention, whereas closed PRs attracted substantially more discussion and exhibited extreme skewness and kurtosis. This indicates that rejected contributions inherently provoke heated debate or prolonged coordination, regardless of a project's overall health. The crucial diagnostic difference emerges in the open PR backlog. In the active control group, open PRs exhibit the highest mean engagement (2.74 comments) and the highest variance, showing an active, ongoing deliberative process. By contrast, open PRs at the time of project death in the inactive cohort show the absolute weakest engagement (mean 0.93, median 0.00). Overall, these distributional properties highlight that while most contributions across GitHub attract few comments, a healthy project actively discusses its pending work. In dying projects, the unresolved backlog is met with profound silence. This specific combination reinforces our conclusion that neglect rather than contentious debate is a true marker of project abandonment.
	
	\begin{table*}[!h]
		\centering
		\caption{Descriptive statistics of discussion volume by PR outcome in Inactive vs Active Projects}
		\label{tab:discussion_by_status}
		\begin{tabular}{lrrr|rrr}
			\toprule
			\multirow{2}{*}{\textbf{Statistic}} & \multicolumn{3}{c}{\textbf{Inactive} (N = 1,296,093)} & \multicolumn{3}{c}{\textbf{Active} (N = 2,671,639)} \\
			\cmidrule(lr){2-4} \cmidrule(lr){5-7}
			& Merged & Closed & Open & Merged & Closed & Open \\
			\midrule
			Mean     & 1.65   & 1.87    & 0.93   & 1.33    & 2.30   & 2.74 \\
			Std      & 3.49   & 4.10    & 2.91   & 3.66    & 4.78   & 6.53 \\
			Median   & 1.00   & 1.00    & 0.00   & 0.00    & 1.00   & 1.00 \\
			Min      & 0.00   & 0.00    & 0.00   & 0.00    & 0.00   & 0.00 \\
			Max      & 383    & 624     & 92     & 1,017   & 464    & 541 \\
			Skew     & 10.71  & 20.53   & 9.57   & 24.25   & 10.82  & 19.21 \\
			Kurtosis & 397.62 & 1997.22 & 157.20 & 3620.80 & 433.92 & 1189.05 \\
			\bottomrule
		\end{tabular}
	\end{table*}
	
	\noindent
	Taken together, these comparative descriptive statistics provide a foundational understanding of PR workflows across both active and inactive projects. The consistently high proportion of unlabeled PRs, combined with extended resolution times and elevated discussion on rejected contributions, confirms that workflow friction and labeling informality are universal platform norms rather than unique symptoms of dysfunction. While both cohorts heavily penalize negative sentiment and face higher friction when integrating unlabeled or innovative work, their overall integration efficiency differs drastically. The true diagnostic difference emerges in the open backlog. In dying projects, unresolved work accumulates rapidly and is met with profound silence, whereas active projects maintain efficient integration and active deliberation on pending tasks. These baseline patterns provide important context, motivating our formal hypothesis testing in RQ1, where we examine exactly how work type, friction, and sentiment relate to individual PR outcomes within the inactive cohort.
	
	\subsection{RQ1: Descriptive Dynamics of PR Friction}
	
\begin{figure}[!h]
		\centering
		\includegraphics[width=0.7\linewidth]{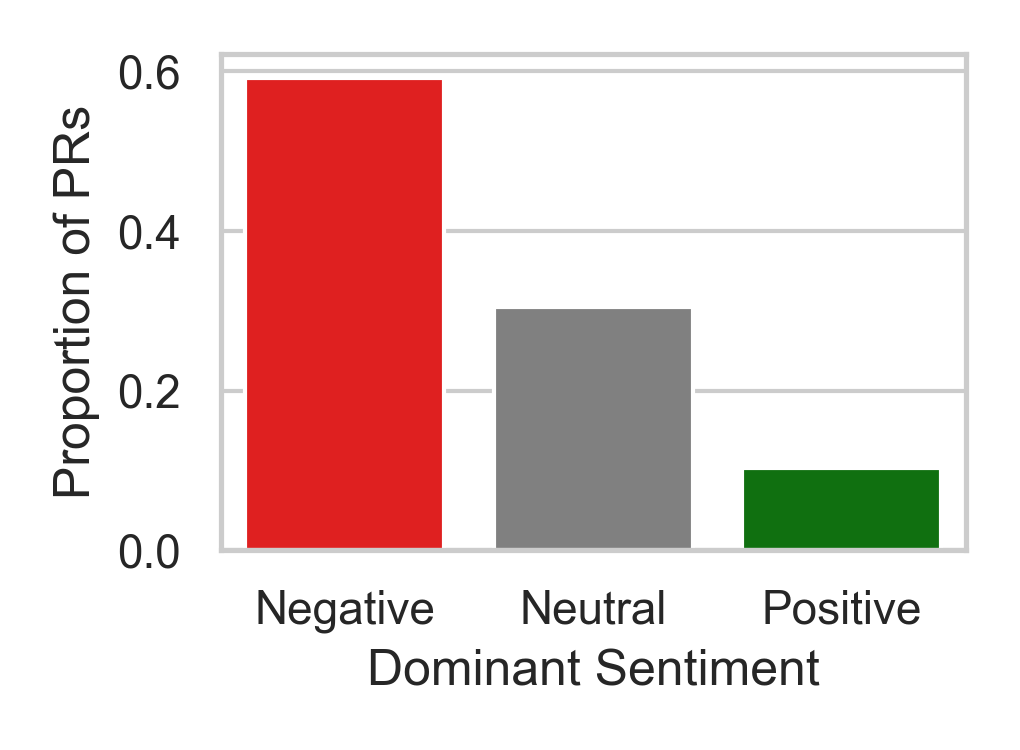}
		\caption{Distribution of PRs by dominant sentiment category.}
		\label{fig:dominant_sentiment}
	\end{figure}

	We begin by examining the distribution of sentiment proportions across the 1.29 million human-driven PRs. 
	At the comment level, negative sentiment was rare (mean = 0.034, median = 0), neutral sentiment was more prevalent (mean = 0.275, median = 0), and positive sentiment was modest (mean = 0.133, median = 0). 
	These skewed distributions confirmed that most PRs contained only a single comment, often neutral or absent, with only a few showing strong emotional tones. 
	When aggregated to the PR level, dominant sentiment categories revealed that 59.2\% of PRs contained at least some negative comments, 30.5\% were dominated by neutral comments, and only 10.3\% were dominated by positive comments (Figure~\ref{fig:dominant_sentiment}). 
	This differs from the comment-level proportions shown earlier in Table~\ref{tab:workflow_metrics}, which were calculated across all comments rather than focusing on the most prevalent sentiment within each PR. This distinction is important because, while individual comments were often neutral, many PRs contained at least one negative remark. As a result, negativity became the most common dominant category.
	
	We next tested the hypotheses proposed in Section~\ref{subsubsec:RQ1}. 
	H1a predicted that innovation PRs would take longer to merge than maintenance PRs. 
	The Kruskal–Wallis test confirmed a statistically significant difference ($H=1145.3$, $p<10^{-250}$), though the effect size was small ($\delta=0.168$). 
	This indicates that innovation PRs were indeed slower to integrate, but the magnitude of difference was small.
	
	H1b predicted that closed PRs would attract more discussion than merged PRs. 
	The Mann–Whitney U test supported this ($U=1.47\times10^{11}$, $p=2.39\times10^{-28}$), but the effect size was negligible ($\delta=0.012$). This is likely due to the fact that most PRs were processed with minimal discussion (see Table~\ref{tab:discussion_by_status}), regardless of their outcome. 
	
	H1c predicted that closed PRs would contain more negative sentiment than merged PRs. 
	The Mann–Whitney U test confirmed this ($U=1.53\times10^{11}$, $p<10^{-300}$), but again with a negligible effect size ($\delta=0.053$). 
	Furthermore, binary toxicity analysis (any negative comment present) revealed that 13.0\% of closed PRs contained negativity compared to 7.7\% of merged PRs ($Z=90.2$, $p<10^{-300}$). 
	Volume-weighted negativity (H1d) produced similar results, with closed PRs exhibiting significantly higher proportions of negative sentiment ($p<10^{-300}$), though the effect size remained negligible ($\delta=0.053$). 
	While the continuous distribution of sentiment proportions showed negligible differences (reflecting the prevalence of neutral comments), the categorical impact was significant: as shown in Table~\ref{tab:sentiment_outcome}, PRs dominated by negativity were rejected at a rate of 36.4\%, compared to just 16.5\% for positive PRs. 
	This confirms that while toxicity is rare in volume, it is fatal when it occurs.
	
	Overall, these hypothesis tests confirm the descriptive dynamics of PR friction within inactive projects. Innovation PRs were slower to merge, closed PRs attracted slightly more discussion and negativity, and binary toxicity was more common in rejections. However, the effect sizes across all tests were small to negligible. When interpreted alongside our active baseline comparison from Section~\ref{subsec:DSRQ1}, this negligible magnitude is expected. Friction and negativity are universal platform norms present even in highly successful projects. Therefore, they are not the primary drivers of PR outcomes or project mortality. This realization motivates our subsequent evolutionary analysis (RQ2), where we shift our focus from these endemic baseline frictions to the true marker of the death spiral.
	
	\begin{tcolorbox}[colback=gray!10,colframe=gray!50,coltext=black, title=RQ1 Takeaway]
		\textbf{Finding:} While formal hypothesis testing confirms that rejected PRs attract slightly more discussion and negativity, the effect sizes are negligible. Combined with our active baseline comparison, this demonstrates that workflow friction and toxicity are universal platform norms. The true differentiator of project mortality is not this day-to-day friction but rather a community's systemic inability to integrate contributions, which ultimately leads to a silent and bloated open backlog.
	\end{tcolorbox}
	
	\subsection{RQ2: Evolutionary Post-Mortem of Workflow}
	
	We next examine how PR workflow dynamics evolved across the lifespan of inactive projects, comparing the first quartile (Q1) to the final quartile (Q4). This analysis tests the "death spiral" pattern: stagnation of innovation, rising merge friction, process collapse, backlog explosion, and social abandonment.

		\textbf{H2a (Stagnation).} Innovation PRs declined modestly over the lifecycle. 
		The Wilcoxon signed-rank test indicated a statistically significant difference 
		($W=151{,}155$, $p=0.016$) at the $\alpha=0.05$ level. Following our Bonferroni correction ($\alpha=0.0071$), this difference is no longer statistically significant. Moreover, the effect size was negligible ($\delta=-0.037$). 
		This suggests that the decline in innovation is inconsistent at adjusted p-values with negligible practical differences.
	Medians in both Q1 and Q4 were zero, reflecting the rarity of innovation PRs overall. Raw counts confirm a modest decline: 10,955 innovation PRs in Q1 versus 10,073 in Q4 (Table~\ref{tab:rq2_counts}). This suggests that innovation did not disappear entirely, but its relative share diminished as projects approached death.
	
\begin{figure}[!h]
		\centering
		\includegraphics[width=0.85\linewidth]{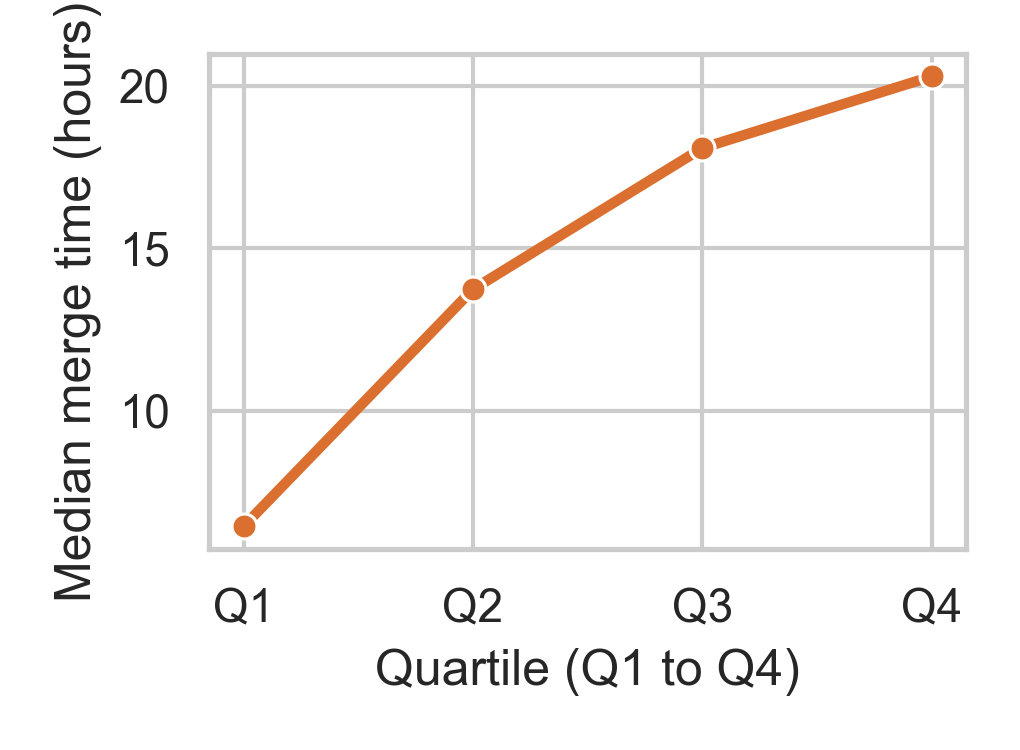}
		\caption{Rising median merge time across quartiles, indicating increasing friction toward project death.}
		\label{fig:friction_spike}
	\end{figure}

	\textbf{H2b (Friction Spike).} As depicted in Figure~\ref{fig:friction_spike}, merge times increased sharply toward project death. Median merge time increased from 7.8 hours in Q1 to 23.2 hours in Q4. The Wilcoxon test confirmed this difference ($W=291{,}761$, $p<10^{-90}$), with a medium effect size ($\delta=0.357$). Aggregate totals show merge effort increased significantly: 30.9 million hours in Q1 versus 65.1 million hours in Q4. This provides strong evidence of escalating friction in the final stages.
	
	\textbf{H2c (Process Collapse).} Closed PR proportions showed no significant change. Median closure rates were 16.2\% in Q1 and 17.5\% in Q4, with the Wilcoxon test non-significant ($p=0.35$) and negligible effect size ($\delta=0.040$). Raw counts remained stable (83,545 closed PRs in Q1 vs. 78,550 in Q4). Thus, rejection did not intensify, but remained a persistent feature across the lifecycle.
	
	\textbf{H2d (Backlog Explosion).} As shown in Figure 3, open PRs accumulated rapidly, following a non-linear pattern. The backlog expanded from 4,491 open PRs in Q1 to 36,641 in Q4, a more than eight-fold increase. The Wilcoxon signed-rank test confirmed this substantial backlog growth ($W=2{,}725.5$, $p<10^{-200}$), with a large effect size ($\delta=0.792$). This significant backlog expansion indicates a clear breakdown in workflow, as unresolved contributions accumulated without maintainer response.
	
	\begin{figure}[!h]
		\centering
		\includegraphics[width=0.85\linewidth]{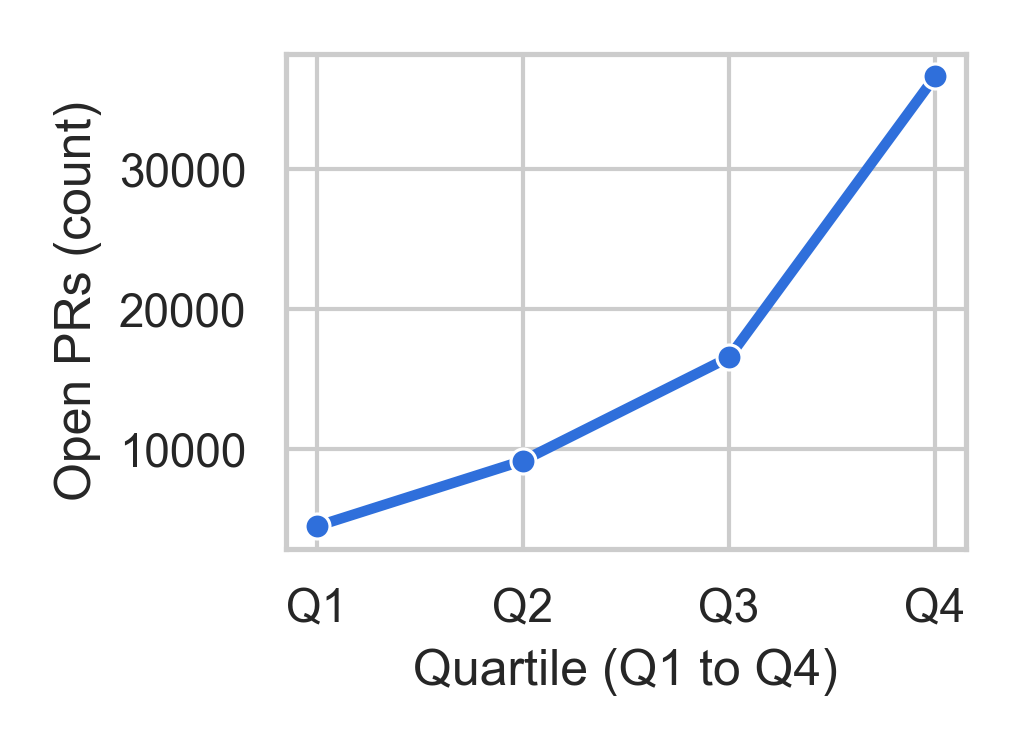}
		\caption{Backlog growth across quartiles, showing exponential accumulation of unresolved PRs.}
		\label{fig:backlog_growth}
	\end{figure}
	
	\textbf{H2e (Social Collapse).} Discussion intensity declined over time. Median discussion volume dropped from 1.0 in Q1 to 0.0 in Q4, while mean values also dropped (0.90 to 0.65). The Wilcoxon test indicated a significant decline ($W=223{,}171$, $p<10^{-18}$), though the effect size was negligible ($\delta=-0.129$). This indicates that PRs in late stages were increasingly ignored rather than debated, reflecting social abandonment. 
	
	To address the possibility that some PRs may have attracted substantive discussion, we analyzed the proportion of PRs with more than two comments. Even among these more engaged PRs, the share declined from 20.9\% in Q1 to 17.3\% in Q4 (Table~\ref{tab:substantive_discussion}). This confirms that substantive engagement also decreased over time, not just median activity.
	
	\begin{table}[!h]
		\centering
		\caption{Proportion of PRs with substantive discussion (>2 comments)}
		\label{tab:substantive_discussion}
		\begin{tabular}{lcc}
			\toprule
			\textbf{Quartile} & \textbf{Total PRs} & \textbf{Substantive PRs (\%)} \\
			\midrule
			Q1 & 322{,}960 & 19.8\% \\
			Q2 & 323{,}793 & 21.0\% \\
			Q3 & 324{,}249 & 20.1\% \\
			Q4 & 325{,}098 & 17.3\% \\
			\bottomrule
		\end{tabular}
	\end{table}
	
	\textbf{H2f (Toxicity Spike).} Negative sentiment proportions declined slightly rather than spiking. Mean negativity decreased from 0.10 in Q1 to 0.07 in Q4, with medians dropping from 0.075 to 0.057. Statistical testing showed no evidence of a spike ($W=688{,}038.5$, $p\approx1.0$, $\delta=-0.094$). This suggests toxicity was not the primary indicator of collapse; instead, silence and neglect defined the final stages.
	
	\textbf{H2g (Labeling formalization).} Unlabeled PRs remained pervasive across all quartiles, with unlabeled proportions consistently high ($\approx80\%$). The Wilcoxon test confirmed no significant increase ($W=515{,}902.5$, $p\approx1.0$, $\delta=-0.079$). This indicates that labeling formalization was endemic from the outset, rather than an outcome of late-stage collapse.
	
	Taken together, these results provide evidence for a ``death spiral'' pattern of workflow collapse. Innovation stagnated, merge friction spiked, and unresolved PRs accumulated into an exponential backlog. Social collapse manifested not in rising toxicity but in declining discussion, as projects were increasingly abandoned by their communities. Labeling formalization was a persistent baseline condition rather than a dynamic driver of decline.
	
	As shown in Table~\ref{tab:rq2_counts}, innovation PRs experienced a modest decline, while closed PR counts decreased. Simultaneously, open PRs (backlog) and merge effort showed a substantial increase, reinforcing the findings of friction and backlog issues.

	\begin{table}[!h]
		\centering
		\caption{Quartile-wise PR workflow metrics (Q1 vs Q4)}
		\label{tab:rq2_counts}
		\begin{tabular}{lcc}
			\toprule
			\textbf{Metric} & \textbf{Q1} & \textbf{Q4} \\
			\midrule
			Innovation PRs (count) & 10,955 & 10,073 \\
			Closed PRs (count)     & 83,545 & 78,550 \\
			Open PRs (backlog)     & 4,491  & 36,641 \\
			Total merge time (hours) & 30.9M & 65.1M \\
			Median merge time (hours) & 7.8 & 23.2 \\
			\bottomrule
		\end{tabular}
	\end{table}
	
	% RQ2 Takeaway
	\begin{tcolorbox}[colback=gray!10,colframe=gray!50,coltext=black,title=RQ2 Takeaway]
		\textbf{Finding:} The ``Death Spiral" is characterized by a rapid, non-linear accumulation of the PR backlog and a spike in merge friction. Rather than rising toxicity, we observe a slight trend toward silence and disengagement, though the effect size of this decline in discussion is statistically negligible.
		
	\end{tcolorbox}
	
	\subsection{RQ3: Explanatory Modeling of Lifespan}
	
	We next examine which PR workflow attributes predict total project lifespan, controlling for license type, programming language, and project size. Table~\ref{tab:rq3_rawstats} reports descriptive statistics for the continuous predictors prior to transformation. Several variables exhibit extreme skewness and kurtosis, such as \textit{stargazers}, \textit{size}, and \textit{median\_merge\_time}, which motivates the use of log transformations in subsequent regression models.
	
	\begin{table*}[!h]
		\centering
		\caption{Descriptive statistics of continuous predictors (before transformation)}
		\label{tab:rq3_rawstats}
		\begin{tabular}{lrrrrrrr}
			\toprule
			\textbf{Variable} & \textbf{Mean} & \textbf{Std} & \textbf{Min} & \textbf{Median} & \textbf{Max} & \textbf{Skew} & \textbf{Kurtosis} \\
			\midrule
			LifeSpan (days)        & 2269.40 & 1286.50 & 4.00 & 2308.00 & 5755.00 & 0.10 & -0.73 \\
			Stargazers (count)     & 4517.94 & 11838.95 & 10.00 & 1440.50 & 325{,}387.00 & 14.50 & 337.48 \\
			Size (KB)              & 105{,}744.23 & 672{,}596.02 & 128.00 & 13{,}747.50 & 24{,}639{,}230.00 & 29.41 & 1040.73 \\
			Median merge time (h)  & 40.34 & 115.19 & 0.002 & 15.95 & 2887.01 & 13.32 & 262.53 \\
			Prop. closed           & 0.21 & 0.13 & 0.00 & 0.18 & 0.95 & 1.80 & 4.98 \\
			Prop. unlabeled       & 0.81 & 0.27 & 0.00 & 0.94 & 1.00 & -1.50 & 1.06 \\
			Avg. prop. negative    & 0.04 & 0.03 & 0.00 & 0.03 & 0.26 & 1.35 & 4.98 \\
			Prop. innovation       & 0.03 & 0.09 & 0.00 & 0.00 & 0.83 & 4.94 & 30.73 \\
			\bottomrule
		\end{tabular}
	\end{table*}
	
	\noindent
	\textit{LifeSpan} represents the total duration (in days) from project creation to its last commit. Its distribution is relatively symmetric, suggesting no transformation is required. \textit{Stargazers} (community popularity) and \textit{Size} (repository volume in KB) are extremely right-skewed, reflecting the presence of a few very large or popular projects; both are log-transformed in regression. \textit{Median merge time} captures the typical latency of PR integration, also highly skewed, and is log-transformed. \textit{Prop. closed} measures the proportion of PRs rejected, bounded between 0 and 1, with moderate skew. \textit{Prop. unlabeled} quantifies labeling formalization as the proportion of PRs without labels. Its high mean (0.81) indicates that most projects neglected systematic labeling. \textit{Avg. prop. negative} reflects the average share of negative sentiment in PR discussions, which remains low overall (mean 0.04). Finally, \textit{Prop. innovation} measures the share of PRs classified as innovation-related. Its skewness shows that most projects contributed little innovation, with a few exceptions.
	
	License categories were grouped into permissive, copyleft, and other. As shown in Table~\ref{tab:rq3_license}, permissive licenses dominate the cohort (64.6\%), followed by other/uncategorized (24.8\%) and copyleft (10.6\%).
	
	\begin{table}[!h]
		\centering
		\caption{Distribution of license categories}
		\label{tab:rq3_license}
		\begin{tabular}{lc}
			\toprule
			\textbf{License Category} & \textbf{Proportion (\%)} \\
			\midrule
			Permissive & 64.6 \\
			Copyleft   & 10.6 \\
			Other      & 24.8 \\
			\bottomrule
		\end{tabular}
	\end{table}
	
	Languages were categorized into the nine most frequent categories, with all others grouped as ``Other.'' Table~\ref{tab:rq3_language} shows that JavaScript (21.1\%) and Python (19.1\%) are most prevalent, followed by TypeScript (9.7\%), C++ (8.3\%), and Go (6.2\%). 
	Together, the top nine languages account for more than 80\% of the sample.
	
	\begin{table}[!h]
		\centering
		\caption{Distribution of primary languages}
		\label{tab:rq3_language}
		\begin{tabular}{lc}
			\toprule
			\textbf{Language} & \textbf{Proportion (\%)} \\
			\midrule
			JavaScript       & 21.1 \\
			Python           & 19.1 \\
			TypeScript       & 9.7 \\
			C++              & 8.3 \\
			Go               & 6.2 \\
			Java             & 5.7 \\
			PHP              & 5.3 \\
			Ruby             & 3.9 \\
			Rust             & 2.8 \\
			Other            & 17.9 \\
			\bottomrule
		\end{tabular}
	\end{table}

	These descriptive results highlight the need for log transformation of highly skewed variables (\textit{stargazers}, \textit{size}, \textit{median merge time}) and justify categorical controls for license and language. They also reveal that labeling formalization (unlabeled PRs) was endemic, innovation contributions were rare, and negativity levels were low. Prior to regression modeling, we analyze multicollinearity using Variance Inflation Factor (VIF).
	Table~\ref{tab:rq3_vif} reports the VIF values for all predictors. 
	A common threshold is VIF $> 10$ as indicative of problematic collinearity. All predictors are well below the threshold, with most values ranging between 1.0 and 3.0, indicating that multicollinearity is not a significant concern.
	
	\begin{table}[!h]
		\centering
		\caption{Variance Inflation Factors (VIF) for regression predictors}
		\label{tab:rq3_vif}
		\begin{tabular}{lc}
			\toprule
			\textbf{Feature} & \textbf{VIF} \\
			\midrule
			C(license\_category)[T.other]        & 2.58 \\
			C(license\_category)[T.permissive]   & 2.69 \\
			C(language\_bin)[T.Go]               & 1.70 \\
			C(language\_bin)[T.Java]             & 1.61 \\
			C(language\_bin)[T.JavaScript]       & 2.90 \\
			C(language\_bin)[T.Other]            & 2.73 \\
			C(language\_bin)[T.PHP]              & 1.62 \\
			C(language\_bin)[T.Python]           & 2.71 \\
			C(language\_bin)[T.Ruby]             & 1.47 \\
			C(language\_bin)[T.Rust]             & 1.34 \\
			C(language\_bin)[T.TypeScript]       & 2.08 \\
			Median merge time                    & 1.05 \\
			Prop. closed                         & 1.17 \\
			Prop. unlabeled                     & 1.42 \\
			Avg. prop. negative                  & 1.24 \\
			Prop. innovation                     & 1.32 \\
			Log stargazers                       & 1.34 \\
			Log size                             & 1.08 \\
			\bottomrule
		\end{tabular}
	\end{table}
	
	\noindent
	
	Following transformation and VIF evaluation, we proceed to estimate the OLS regression model to predict project lifespan, testing hypotheses H3a through H3d.
	
	We initially estimated the OLS regression with conventional standard errors. 
	Residual diagnostics (Omnibus $p=0.001$, Jarque--Bera $p=0.0036$) indicated mild deviations from normality and potential heteroscedasticity\footnote{Heteroscedasticity refers to a condition in regression analysis where the variability of the error term is not constant across all levels of the independent variables.}. 
	Although OLS coefficients remain unbiased under heteroscedasticity, conventional standard errors can lead to incorrect inference. 
	To address this, we re-estimated the model using heteroscedasticity-consistent robust standard errors (HC3). HC3 is recommended in finite samples because it adjusts more strongly for leverage points and yields more reliable inference than HC0 or HC1. Monte Carlo evidence confirms that HC3 performs well in these situations~\cite{long2000hc3}. Accordingly, all regression results reported below use HC3 robust standard errors.
	
	\noindent
	Table~\ref{tab:rq3_regression} reports the HC3-robust OLS regression results predicting project lifespan. The model explains 38.3\% of the variance in lifespan ($R^2 = 0.383$). 
	Residual diagnostics indicated mild heteroscedasticity, which motivated the use of HC3 robust standard errors \citep{long2000hc3}.

	\begin{table*}[!h]
		\centering
		\caption{OLS regression predicting project lifespan (HC3 robust standard errors)}
		\label{tab:rq3_regression}
		\begin{tabular}{lrrrrr}
			\toprule
			\textbf{Predictor} & \textbf{Coef.} & \textbf{Robust SE (HC3)} & \textbf{z} & \textbf{p-value} & \textbf{95\% CI} \\
			\midrule
			log Median merge time & 71.55 & 21.66 & 3.30 & 0.001 & [29.09, 114.01] \\
			Prop. closed          & 194.32 & 205.00 & 0.95 & 0.343 & [-207.47, 596.11] \\
			Prop. unlabeled      & 1158.57 & 100.57 & 11.52 & 0.000 & [961.45, 1355.68] \\
			Avg. prop. negative   & 8862.59 & 1151.47 & 7.70 & 0.000 & [6605.75, 11119.43] \\
			Prop. innovation      & 1150.83 & 296.09 & 3.89 & 0.000 & [570.51, 1731.14] \\
			log Stargazers        & 202.66 & 14.48 & 13.99 & 0.000 & [174.28, 231.04] \\
			log Size              & -21.33 & 12.99 & -1.64 & 0.101 & [-46.80, 4.14] \\
			\bottomrule
		\end{tabular}
	\end{table*}
	
	\noindent
	\textbf{Hypothesis tests.} 
	\begin{itemize}
		\item \textbf{H3a (Friction)}: Contrary to expectation, higher median merge time is a \emph{positive} predictor of lifespan ($\beta = 71.6$, $p = 0.001$). This suggests that longer merge latency is associated with longer-lived projects, consistent with the idea that complexity and deliberation contribute to longevity rather than accelerating failure.
		
		\item \textbf{H3b (Waste)}: The proportion of closed PRs is not significant ($p = 0.343$), indicating that rejection rates do not predict lifespan.
		\item \textbf{H3c (Labeling formalization)}: Labeling formalization (unlabeled PRs) is a strong positive predictor ($\beta = 1158.6$, $p < 0.001$). This counterintuitive result suggests that labeling discipline is not a determinant of survival; rather, projects with less formalization persisted longer, possibly reflecting informal governance structures.
		\item \textbf{H3d (Toxicity)}: Average negativity is a strong positive predictor ($\beta = 8862.6$, $p < 0.001$). This indicates that negativity is more a byproduct of longevity than a cause of mortality, aligning with our descriptive findings.
		\item \textbf{H3e (Innovation)}: Innovation proportion is a significant positive predictor of lifespan ($\beta = 1240.3$, $p < 0.001$). This supports the hypothesis that innovation sustains survival, particularly in longer-lived projects where value-adding contributions remain central.
	\end{itemize}
	
	Community popularity (log stargazers) is a highly significant positive predictor ($\beta = 202.7$, $p < 0.001$), while project size is not significant. 
	Language effects are notable: PHP and Ruby projects show significantly longer lifespans, while Rust and TypeScript projects show shorter lifespans. 
	License type has limited effect, with only ``Other'' licenses marginally predicting longer lifespan.
	
	To further investigate heterogeneity across the lifespan distribution, we estimated quantile regressions at the 25th, 50th, and 75th percentiles. 
	These results reveal important differences between shorter- and longer-lived projects. 
	At the 25th percentile, labeling formalization and negativity remain strong positive predictors, while innovation is not significant. 
	At the median, innovation (H3e) emerges as a significant positive predictor alongside labeling formalization, negativity, and popularity. 
	At the 75th percentile, the effects of labeling formalization and negativity intensify, innovation becomes highly significant, and project size turns negative, suggesting maintenance burden in very long-lived projects. 
	Median merge time remains a positive predictor across all quantiles, rejecting H3a, while rejection rates remain non-significant, refuting H3b. 
	Language effects also vary: PHP and Ruby consistently predict longer lifespans, while Rust and TypeScript predict shorter lifespans, particularly in median and upper quantiles.
	
	\noindent
	Taken together, the quantile regressions reinforce and extend the OLS findings. 
	H3a (friction) is refuted: longer merge times are associated with longer lifespans across all quantiles. 
	H3b (waste) is refuted: rejection rates do not predict lifespan at any quantile. 
	H3c (Labeling formalization) is refuted: labeling formalization consistently predicts longer lifespan, with stronger effects in longer-lived projects. 
	H3d (toxicity) is refuted: negativity is a robust positive predictor across all quantiles, indicating that negative sentiment accumulates in projects that persist rather than driving early mortality. 
	H3e (innovation) is supported: innovation proportion predicts longer lifespan, particularly in median and upper quantiles, highlighting its role in sustaining survival.
	Popularity (log stargazers) is a consistently strong positive predictor, while project size only becomes a negative factor in the longest-lived projects, reflecting maintenance burdens.
	
	\noindent
	Overall, these results demonstrate that OSS mortality is not explained by workflow discipline or sentiment dynamics. 
	Instead, abandonment patterns and ecosystem value dominate survival outcomes. 
	Friction, label formalization, and negativity scale with longevity rather than accelerating failure, while innovation and popularity sustain survival in longer-lived projects. 
	This quantile perspective highlights that predictors of lifespan behave differently across shorter- and longer-lived projects, underscoring the importance of ecosystem dynamics over PR-level workflow efficiency in shaping OSS project mortality.
	
	\subsection*{Sensitivity Analysis: Alternative Inactivity Thresholds}\label{subsec:sensitivity_analysis}
	
	A common critique in OSS mortality research is that the definition of ``inactive" projects is inherently debatable. 
	Previous studies have used thresholds between 6 months and over a year, without reaching a consensus, and the choice of a single cutoff point influences both dataset composition and results. 
	To address this concern, we conducted robustness checks using alternative inactivity thresholds of 9 months (270 days) and 12 months (365 days). 
	Tables~\ref{tab:rq3_robust9} and \ref{tab:rq3_robust12} report the HC3-robust OLS regression results under these definitions.
	
	\begin{table*}[!h]
		\centering
		\caption{Robustness Check: OLS regression predicting project lifespan (Inactivity $\geq$ 9 months, HC3 robust SEs)}
		\label{tab:rq3_robust9}
		\begin{tabular}{lrrrrr}
			\toprule
			\textbf{Predictor} & \textbf{Coef.} & \textbf{Robust SE (HC3)} & \textbf{z} & \textbf{p-value} & \textbf{95\% CI} \\
			\midrule
			log Median merge time & 70.58 & 26.97 & 2.62 & 0.009 & [17.72, 123.44] \\
			Prop. closed          & 241.29 & 266.23 & 0.91 & 0.365 & [-280.51, 763.09] \\
			Prop. unlabeled      & 1075.03 & 127.77 & 8.41 & 0.000 & [824.61, 1325.45] \\
			Avg. prop. negative   & 7833.22 & 1350.49 & 5.80 & 0.000 & [5186.31, 10480.95] \\
			Prop. innovation      & 1347.13 & 378.43 & 3.56 & 0.000 & [605.42, 2088.83] \\
			log Stargazers        & 211.15 & 17.68 & 11.94 & 0.000 & [176.50, 245.80] \\
			log Size              & -9.48 & 16.12 & -0.59 & 0.556 & [-41.08, 22.12] \\
			\bottomrule
		\end{tabular}
	\end{table*}
	
	\begin{table*}[!h]
		\centering
		\caption{Robustness Check: OLS regression predicting project lifespan (Inactivity $\geq$ 12 months, HC3 robust SEs)}
		\label{tab:rq3_robust12}
		\begin{tabular}{lrrrrr}
			\toprule
			\textbf{Predictor} & \textbf{Coef.} & \textbf{Robust SE (HC3)} & \textbf{z} & \textbf{p-value} & \textbf{95\% CI} \\
			\midrule
			log Median merge time & 61.57 & 29.35 & 2.10 & 0.036 & [4.04, 119.10] \\
			Prop. closed          & 624.13 & 303.77 & 2.06 & 0.040 & [28.76, 1219.50] \\
			Prop. unlabeled      & 923.80 & 143.94 & 6.42 & 0.000 & [641.69, 1205.91] \\
			Avg. prop. negative   & 6837.71 & 1486.54 & 4.60 & 0.000 & [3924.14, 9751.27] \\
			Prop. innovation      & 930.42 & 357.58 & 2.60 & 0.009 & [229.58, 1631.25] \\
			log Stargazers        & 234.60 & 20.06 & 11.69 & 0.000 & [195.27, 273.92] \\
			log Size              & -25.52 & 17.97 & -1.42 & 0.156 & [-60.75, 9.71] \\
			\bottomrule
		\end{tabular}
	\end{table*}

	\noindent
	The results are highly consistent with our main specification (6 months). 
	At 9 months, the model explains 38.3\% of the variance in lifespan ($R^2 = 0.383$), with proportion unlabeled, negativity, innovation, and popularity remaining strong positive predictors. 
	At 12 months, explanatory power increases slightly ($R^2 = 0.434$), and the same predictors remain significant. 
	Median merge time continues to be a positive predictor across thresholds, refuting H3a. 
	Rejection rates remain non-significant at 9 months, but become marginally significant at 12 months ($\beta = 624.1$, $p = 0.040$), suggesting that waste may play a role only in the longest inactivity definition. 
	Labeling formalization (H3c) and toxicity (H3d) are consistently refuted, as both remain strong positive predictors across thresholds. Innovation (H3e) is consistently supported across thresholds, confirming its role as a robust positive predictor of lifespan. Popularity sustains survival in longer-lived projects, while project size remains non-significant or weakly negative.
	
	\noindent
	Taken together, these robustness checks confirm that our explanatory findings are not artifacts of the 6-month inactivity definition. 
	Regardless of whether inactivity is defined as 6, 9, or 12 months, the same core predictors emerge: proportion unlabeled, negativity, innovation, and popularity are associated with longer lifespans, while friction and rejection rates do not drive mortality. This strengthens the validity of our conclusion that OSS mortality is shaped by abandonment and ecosystem dynamics rather than PR-level workflow discipline.

	% RQ3 Takeaway
	\begin{tcolorbox}[colback=gray!10,colframe=gray!50,coltext=black,title=RQ3 Takeaway]
		\textbf{Finding:} Project lifespan is predicted by ecosystem value (Popularity, Innovation) rather than workflow efficiency. Friction and labeling formalization are positive predictors of longevity, suggesting they are byproducts of survival rather than drivers of death.
	\end{tcolorbox}
	
	\section{Case Studies}\label{sec:casestudy}
\begin{figure*}
		\centering
		\includegraphics[width=\textwidth]{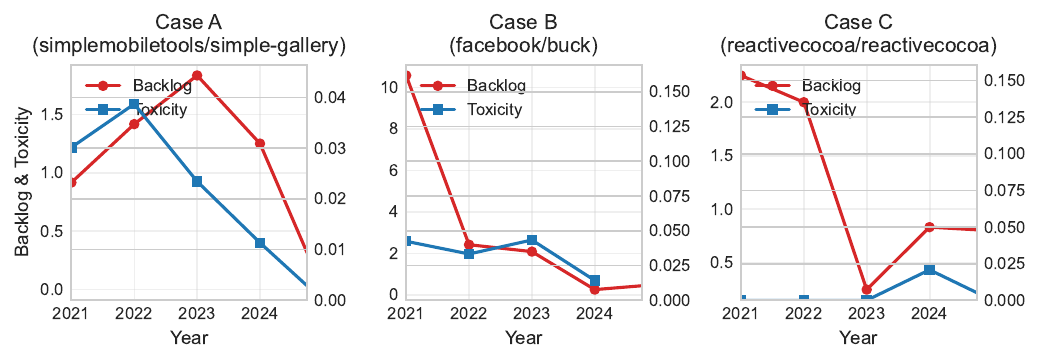}
		\caption{Backlog and toxicity trends for three case study repositories (Simple-Gallery, Facebook Buck, ReactiveCocoa). 
			The red line shows the yearly average backlog, computed as the cumulative count of open PRs (creation events $+1$, resolution events $-1$) resampled to yearly frequency. The blue line shows toxicity, calculated as the proportion of negative comments (\texttt{prop\_negative}) across PRs, resampled to yearly frequency and anchored at zero.}
		
		\label{fig:backlogtrends}
	\end{figure*}

	To complement our statistical findings and provide concrete examples, we analyzed three representative projects from our inactive cohort. These projects were selected after a thorough review of the inactive project list. We chose them to illustrate diverse modes of project death. Figure~\ref{fig:backlogtrends} shows their backlog and toxicity trends over the last three years.
	
	\subsection{Simple-Gallery}
	Simple-Gallery\footnote{\url{https://github.com/SimpleMobileTools/Simple-Gallery}} was a highly popular open-source Android application, valued for its simplicity and privacy features. For several years, contributions flowed steadily, with the backlog averaging around 1-2 PRs between 2020 and 2022. However, following the project's acquisition by a commercial entity in late 2023, the community dynamics shifted drastically. By 2024, the PR average backlog collapsed completely to 0.00. The last commit was recorded on June 11, 2024, marking the definitive end of development activity. On the surface, the repository still appeared active. The README was updated with extensive donation links (IBAN, Bitcoin, Ethereum, Patreon, PayPal) and a badge to ``Get it on F-Droid." These signals suggested continuity, but they are misleading. No code was being integrated, and no workflow remained. Meanwhile, the issue tracker revealed the real situation. By late 2025, there were 377 open issues and 1,457 closed, with the last closure on May 15, 2024. Users reported critical failures such as ``Please update for new API" and ``App no longer works." None of these were addressed. Instead, community members began redirecting frustrated users to a fork: ``Go to Fossify page\footnote{\url{https://github.com/FossifyOrg}} and you will understand." This confirms that the community did not simply dissolve; they actively migrated their labor to Fossify, a fork established to preserve the project's open-source integrity. This case align with our findings.
	
\begin{itemize}
		\item \textbf{RQ1 (Friction and Negativity):} The issue tracker shows complaints and frustration, but these did not escalate into lasting toxicity within the PR workflow. Instead, the collapse was marked by a clear ``Contribution Gap."
		\item \textbf{RQ2 (Evolutionary Death Spiral):} The backlog collapse and the stop in commits confirm the common pattern of workflow decline, marked by silence and disengagement rather than rising toxicity.
		\item \textbf{RQ3 (Explanatory Modeling):} This case shows that workflow efficiency was not the deciding factor. What mattered was \textit{Ecosystem Value} and \textit{Governance Signals}. Once the governance model shifted (through acquisition) and trust broke, contributors moved their efforts elsewhere.
	\end{itemize}
	
\subsection{Facebook Buck}
	Facebook Buck\footnote{\url{https://github.com/facebook/buck}} was a widely used build system at Meta. Unlike the chaotic collapse of Simple-Gallery, this project followed a strategic shutdown. As Buck2 replaced it internally, the average backlog was systematically reduced: from a high of 11 PRs in 2020 down to 0 in 2023. Half-yearly data shows the same pattern, with backlog values dropping to zero in 2023Q1 and 2023Q3. The last commit was made on April 19, 2023, updating the README to point users toward Buck2. Shortly after, the repository was formally archived on November 10, 2023, and is now read-only.
	The signals here were explicit. The README clearly stated: ``This project is archived and will no longer be maintained. We recommend using Buck2." The issue tracker also reflects this transition: 204 open issues and 1,277 closed, with the last closure in November 2022. PR activity tapered as well, with 22 open PRs and 1,168 closed, the last closure recorded in February 2024.	
	This case demonstrates that not all project deaths are chaotic. Buck's decline was strategic and controlled, with pending work resolved before archival. This case aligned to our results as:
	
	\begin{itemize}
		\item \textbf{RQ1 (Friction and Negativity):} The issue tracker shows user questions and frustrations, but these did not escalate into lasting toxicity. The workflow remained orderly, with PRs processed and closed before archival.
		\item \textbf{RQ2 (Evolutionary Death Spiral):} Instead of a backlog explosion, Buck showed backlog cleanup. The decline ended in silence after the wind-down, confirming that the final marker of mortality is the stop in workflow activity.
		\item \textbf{RQ3 (Explanatory Modeling):} This case highlights that workflow efficiency was not the deciding factor. Mortality was driven by \textit{Ecosystem Value} (Buck2 replacing Buck) and clear \textit{Governance Signals} (explicit archival notice), which guided contributors to move on.
	\end{itemize}
	
	\subsection{ReactiveCocoa}	
	ReactiveCocoa~\footnote{\url{https://github.com/ReactiveCocoa/ReactiveCocoa}} was once the leading framework for functional reactive programming on iOS. For years, it provided developers with powerful abstractions and attracted steady contributions. However, the release of Apple's native Combine framework changed the landscape. With a platform-supported alternative available, ReactiveCocoa's relevance declined, and contributors gradually moved away. The workflow signals reflect this shift. The average backlog remained consistently low and flat, around 0.8 PRs in 2023-2024. This happens not because of efficiency, but because of new contributions stopped coming in. Half-yearly data shows the same pattern, with values staying between 0.3 and 1.0. The last commit was recorded on May 19, 2024, adding support for xrOS, but no further development followed. The issue tracker also shows signs of fading engagement. Out of more than 2,000 closed issues, only 4 remain open, mostly related to compatibility with newer Apple requirements. Pull request activity slowed as well. Just 7 open PRs remain, while the last merged PR was in May 2024. Unlike Case B (Buck), there was no explicit archival notice. The project simply became quiet as developers migrated to Apple's native solution. This case shows a different pathway to mortality: ecosystem displacement. ReactiveCocoa did not collapse due to friction or governance failure, nor was it deliberately shut down. Instead, its value proposition evaporated when the ecosystem itself shifted. Even technically sound projects can fade when their platform introduces a native alternative. This case validates our RQ3 finding on ecosystem dynamics. The ``death spiral" here was not marked by conflict or backlog explosion, but by silence and migration. It confirms that mortality can occur quietly when ecosystem relevance disappears.
	
	These cases show that while the reasons for project decline differ. Community exit (Simple-Gallery), planned closure (Buck), or ecosystem shift (ReactiveCocoa), the \textit{outcome} in the PR workflow is the same: silence in contribution. Whether the backlog grows, is reduced, or stays flat, the key signal of mortality is the permanent stop in collaborative activity.
	\section{Discussion}\label{sec:Discussion}
	
	Our findings contribute to the ongoing debate on OSS project mortality by reframing it as a socio-technical phenomenon viewed through the lens of PR workflow dynamics. Prior macro-level research has emphasized developer abandonment, truck factor risks, maintainer burnout, and ecosystem dependencies~\cite{linaaker2024sustaining,nourry2024myth,avelino2019abandonment,kaur2022exploring,coelho2017modern}. In contrast, our post-mortem analysis of 1,736 inactive projects shows that micro-level workflow signals (merge latency, rejection rates, labeling discipline, and sentiment) do not act as causal determinants of failure. These signals scale with longevity and emerge as byproducts of survival. OSS mortality therefore depends on inherent project value and ecosystem dynamics rather than workflow efficiency.
	
	Our descriptive analysis (RQ1) confirms that friction and negativity are pervasive, but comparing them against our active control group reframes their meaning entirely. While innovation PRs take longer to merge (consistent with~\cite{xu2025predicting}) and rejected PRs attract higher discussion volume and negativity~\cite{Asri2019}, these patterns are virtually identical in healthy, active projects. Active projects heavily penalize negative sentiment and experience substantial friction when integrating complex work. Therefore, PR-level friction and negativity are not systemic signals of mortality, but rather universal norms of the GitHub platform. The true diagnostic difference emerges in how the community handles pending work: active projects maintain high deliberation on their open backlogs (averaging 2.74 comments per PR), whereas dying projects meet their unresolved backlogs with profound silence (averaging 0.93 comments).

	The evolutionary analysis (RQ2) characterizes collapse as a gradual shift from endemic friction to total social abandonment, termed the death spiral. The backlog of open pull requests increased by more than eightfold from the first to the final quartile (a statistically Large effect size). Median merge time also rises sharply from 7.8 to 23.2 hours. Furthermore, we observe a slight trend toward decreasing discussion volume in late stages, suggesting neglect rather than active contention. However, as the statistical effect size of this overall decline in discussion is negligible, we cannot classify 'silence' as the primary symptom of collapse. Instead, the defining symptom of the death spiral is the exponential accumulation of the unresolved, completely ignored PR backlog. This finding is supported by our case study of \textit{Simple-Gallery}, where the backlog collapsed because the community quietly moved to a fork. Negative sentiment does not spike toward death (H2f refuted); instead, it declines slightly. Projects die when maintainers withdraw from the human infrastructure~\cite{linaaker2024sustaining} and stop responding, leaving contributions unresolved~\cite{kaur2022exploring,coelho2017modern}. As seen in the \textit{Facebook Buck} case, this withdrawal can also be a planned shutdown rather than a collapse. Distinguishing between such \textit{Strategic Completions} and actual \textit{Abandonments} is an important challenge for mortality metrics.
	
	The explanatory modeling yields several unexpected findings that challenge conventional assumptions about organizational failure. Higher median merge time positively predicts lifespan across all quantiles and robustness checks, refuting H3a. While studies of active projects associate positive sentiment with faster closure~\cite{Asri2019} and rejected PRs with longer discussions~\cite{Golzadeh2019}, aggregate slowness in our analysis signals rigor rather than dysfunction. Longer-lived projects exhibit greater complexity~\cite{park2025analyzing}, and deliberate review reflects high stakes and architectural constraints worthy of investment. Innovation proportion and community popularity (measured as log stargazers) emerge as the strongest positive predictors of lifespan. This result aligns with recent abandonment-prediction models that treat feature-PR ratios as indicators of functional stagnation~\cite{xu2025predicting,lu2025open}. Sustained value-adding contributions provide the primary reason for core developers and the wider community to remain engaged~\cite{park2025analyzing}.
	
	Labeling formalization (high proportion of unlabeled PRs) also positively predicts lifespan, refuting H3c. Informal, low-overhead governance often proves more sustainable than rigid labeling schemes in mid-sized or long-lived communities~\cite{Alami2020}. Similarly, average negativity positively predicts lifespan, refuting H3d. Although negativity harms individual PR outcomes (PRs dominated by negative sentiment close at 36.4\% versus 16.5\% for positive-dominant ones), its accumulation at project level marks enduring, high-impact debates rather than imminent demise. Rejection rates show no consistent predictive power, refuting H3b.
	
	Ultimately, OSS survival hinges on ecosystem value and social capital rather than workflow discipline~\cite{linaaker2024sustaining,park2025analyzing}. The \textit{ReactiveCocoa} case shows this clearly. The project did not fail due to governance or friction, but simply lost relevance when Apple's Combine framework replaced its value. Furthermore, Projects that maintain innovation proportion and popularity endure longest, even amid slow integration and contentious debate. Robustness checks across 9-month and 12-month inactivity thresholds confirm that friction, labeling formalization, and negativity are positively associated with longer lifespans as byproducts of persistence, while innovation and popularity consistently drive survival.
	
	\subsection*{Implications}
	
	The study provides significant implications for maintainers, researchers, and developers by reframing OSS mortality as a socio-technical phenomenon driven by ecosystem value and social abandonment, rather than workflow inefficiency. The post-mortem analysis of PR workflow dynamics highlights that conventional efficiency metrics are misleading indicators of long-term health, and that survival is instead determined by sustained innovation, popularity, and continued human engagement.
	
	For maintainers, the findings suggest a shift in priorities. Sustaining a high innovation proportion ($\text{prop\_innovation}$) should be central, as it consistently predicts project longevity. Longer merge times should be accepted as part of rigorous review in complex projects, rather than treated as failure signals. Informal governance practices, such as tolerating unlabeled PRs, can be viable in sustaining collaboration, and occasional negativity in discussions should be understood as part of high-stakes debates. The true warning signs are disengagement and neglect: a sharp drop in discussion activity and the exponential growth of unresolved PR backlogs. Monitoring these signals can help maintainers intervene before collapse.
	
	For researchers, the study calls for a reframing of mortality models. Indicators such as friction, labeling formalization, rejection rates, and negativity should be treated as correlates of persistence rather than causal drivers of failure. Future predictive models should emphasize human engagement metrics, particularly stagnation of innovation and declining discussion volume, as leading indicators of collapse. Extending this framework to longitudinal studies of active projects will allow validation of the death spiral mechanism in real time. Qualitative work with maintainers can further clarify the role of informal governance and the meaning of sentiment in sustaining communities. Tool developers should also design metrics that capture micro-level signals of abandonment, rather than relying solely on commit activity.
	
	For developers and users, the findings provide guidance in evaluating project viability. Projects with strong community popularity ($\text{log\_stargazers}$) and sustained innovation are more likely to endure. Contributors should not interpret slow merges or contentious discussions as signs of decline; these often accompany long-lived, complex projects. Instead, silence is the clearest warning sign: when contributions receive little or no discussion, it signals social abandonment and impending collapse. Users and contributors should prioritize engagement with projects that demonstrate ongoing innovation and active dialogue.
	
	\section{Threats to Validity}\label{sec:Threats}
	
	The study acknowledges several limitations and potential threats to validity. These are organized into the standard categories of internal, construct, external, and conclusion validity.
	
	\subsection{Internal Validity}
	Internal validity concerns whether the observed relationships are genuine or confounded by other factors. 
	We relied on a heuristic keyword-based approach (e.g., ``bot'' in login) combined with manual validation. While more advanced tools may capture additional automated accounts, our manual verification of high-volume contributors reduced the risk of retaining significant automated noise. We prioritized precision (avoiding the removal of human contributors) over recall. 
	Our OLS regression identifies predictors of lifespan but cannot establish causality. For example, while we find that labeling formalization predicts longevity, this is an association rather than a recommendation to stop labeling PRs.
	
	\subsection{Construct Validity}
	Construct validity concerns whether the measures accurately capture the intended concepts. 
	We operationalized labeling formalization as the proportion of unlabeled PRs. This metric captures \textit{on-platform} governance discipline rather than total organizational maturity, since many projects may use external tools for workflow management. The fact that a majority of active projects also lack labels suggests this is a platform-wide norm rather than a unique characteristic of dying projects.
	We defined mortality as a six-month halt in commits. This definition may conflate failed projects with completed ones. To address this, we distinguished archetypes in our Case Studies (Section~\ref{sec:casestudy}), showing that while the intent differs (community exit vs.\ strategic closure), the workflow signal (cessation of activity) is identical. Additionally, we treated closed PRs as rejected. However, a PR may also be closed due to duplication or externally managed contributions.
	
	\subsection{External Validity}
	External validity concerns the generalizability of findings beyond the studied sample. 
	Our evolutionary analysis focused on inactive projects. To provide context, we introduced a 1:1 structurally matched control group of 1,736 active projects in Section~\ref{subsec:4.1}. This allows comparison, though our ``Death Spiral" model applies specifically to the mortality phase.
	Our dataset covers project histories up to September 2024. While this captures a broad history, newer shifts in tooling (e.g., AI-generated PRs) may alter future mortality dynamics.
	This study is restricted to GitHub. Ecosystems with different governance structures (e.g., Apache Foundation projects using mailing lists) may exhibit different mortality signals.
	
	\subsection{Conclusion Validity}
	Conclusion validity concerns whether the statistical conclusions are sound and supported by the data. 
	The use of OLS regression with robust HC3 standard errors provided interpretable results, but inference depends on model assumptions. While robustness checks across alternative inactivity thresholds strengthen confidence in the findings, reliance on a single modeling approach may limit the breadth of conclusions. Additionally, the exclusion of commit-level and issue-level dynamics means that explanatory models capture only PR workflow attributes, potentially omitting other relevant predictors of project survival.
	
	\section{Conclusion and Future Work}\label{sec:Conclusion}
	
	This study provides the first large-scale post-mortem analysis of PR workflow dynamics, reframing Open Source mortality as a socio-technical phenomenon driven by abandonment and ecosystem value rather than workflow efficiency.
	
	By comparing 1,736 inactive repositories against a 1:1 structurally matched control group of 1,736 active projects, we dismantled the ``Friction Hypothesis." We found that friction (slower merges, unlabeled PRs) is endemic to the platform, not a unique killer. Survivors are often slower than dying projects, suggesting that deliberation aids longevity.
	
	Our evolutionary analysis identified a ``Death Spiral" marked largely by silence. As illustrated in our qualitative case studies, ranging from the ``Mutiny" of Simple-Gallery to the ``Strategic Completion" of Facebook Buck, the universal signal of mortality is not toxicity, but the cessation of human interaction. The backlog explodes not because maintainers fight, but because they leave.
	
	Collapse is ultimately defined by social abandonment. Because the overall decline in discussion volume has a statistically negligible effect size, the clearest and most robust signal of the death spiral is the rapid, non-linear accumulation of an unresolved, ignored PR backlog. Project survival is determined by sustained innovation and community engagement, while friction, labeling formalization, and negativity scale with longevity as byproducts of persistence rather than causes of failure.
	
	Future work should extend this ``autopsy" approach to real-time intervention tools. Since silence (falling discussion volume) is a leading indicator of collapse. While future bots could detect this signal early, such interventions may not always alter a project trajectory. As demonstrated in our case studies, mortality may be driven by inevitable external ecosystem shifts. Beyond this, longitudinal studies of active projects could validate the death spiral mechanism in real time. Integrating commit-level and issue-level dynamics alongside PR workflows would provide a more holistic view of project decline. Qualitative studies with maintainers could enrich quantitative findings, particularly regarding informal governance and the meaning of negative sentiment. Finally, replication across other ecosystems such as GitLab and Bitbucket will be essential to test the generalizability of the death spiral mechanism beyond GitHub.

\bibliographystyle{elsarticle-num}
\bibliography{references_r2}

\end{document}